\documentclass[twocolumn, resetfootnote]{aastex702} %
\usepackage{amsmath}
\usepackage{natbib}
\usepackage{float}
\usepackage{CJKutf8} %
\usepackage{placeins}
\usepackage{subfigure}
\usepackage{graphicx}

\newcommand{\teff}{$T_\mathrm{eff}$}
\newcommand{\logg}{$\log{g}$}

\newcommand{\vsini}{$v\sin{i}$}
\newcommand{\kms}{km~s$^{-1}$}

\newcommand{\msun}{M$_{\odot}$}

\submitjournal{ApJ}

\shorttitle{Metal-poor Brown Dwarf Kinematics}
\shortauthors{Hsu et al.}

\graphicspath{{./}}

\begin{document}

\title{Metal-poor Brown Dwarf Kinematics from JWST NIRSpec Spectroscopy}

\correspondingauthor{Chih-Chun Hsu}
\email{chsu@northwestern.edu}

\author[0000-0002-5370-7494]{Chih-Chun Hsu}
\affil{Center for Interdisciplinary Exploration and Research in Astrophysics (CIERA), Northwestern University,
1800 Sherman Ave, Evanston, IL, 60201, USA}
\email{chsu@northwestern.edu}

\author[0000-0002-9807-5435]{Christopher A. Theissen}
\affiliation{Department of Astronomy \& Astrophysics, University of California San Diego, La Jolla, CA 92093, USA}
\email{ctheissen@ucsd.edu}

\author[0000-0002-6523-9536]{Adam J.\ Burgasser} %
\affiliation{Department of Astronomy \& Astrophysics, University of California San Diego, La Jolla, CA 92093, USA}
\email{ctheissen@ucsd.edu}

\author[0000-0002-6294-5937]{Adam C. Schneider}
\affil{United States Naval Observatory, Flagstaff Station, 10391 West Naval Observatory Rd., Flagstaff, AZ 86005, USA}
\email{aschneid10@gmail.com}

\author[0000-0003-2094-9128]{Christian Aganze}
\affiliation{Department of Physics, Stanford University, Stanford, CA 94305, USA}
\email{caganze@ucsd.edu}

\author[0000-0002-1125-7384]{Aaron M. Meisner}
\affiliation{NSF National Optical-Infrared Astronomy Research Laboratory, 950 N. Cherry Ave., Tucson, AZ 85719, USA}
\affiliation{Center for Astrophysics $|$ Harvard \& Smithsonian, 60 Garden St., Cambridge, MA 02138, USA}
\affiliation{Radcliffe Institute for Advanced Study at Harvard University, 10 Garden Street, Cambridge, MA 02138, USA}
\email{aaron.m.meisner@gmail.com}

\author[0000-0002-2592-9612]{Jonathan Gagn\'e}
\affiliation{Plan\'etarium de Montr\'eal, Espace pour la Vie, 4801 av. Pierre-de Coubertin, Montr\'eal, Qu\'ebec, Canada}
\affiliation{Trottier Institute for Research on Exoplanets, Universit\'e de Montr\'eal, D\'epartement de Physique, C.P.~6128 Succ. Centre-ville, Montr\'eal, QC H3C~3J7, Canada}
\email{jonathan.gagne@montreal.ca}

\author[0000-0002-3612-8968]{Nicolas Lodieu}
\affiliation{Instituto de Astrof\'isica de Canarias (IAC), Calle V\'ia L\'actea s/n, E-38200 La Laguna, Tenerife, Spain}
\affiliation{Departamento de Astrof\'isica, Universidad de La Laguna (ULL), E-38206 La Laguna, Tenerife, Spain}
\email{nlodieu@iac.es}

\author[0000-0003-3047-607X]{ZengHua Zhang}
\affiliation{School of Astronomy and Space Science, Nanjing University, 163 Xianlin Avenue, Nanjing 210023, China}
\affiliation{Key Laboratory of Modern Astronomy and Astrophysics, Nanjing University, Ministry of Education, Nanjing 210023, China}
\email{zenghuazhang@hotmail.com}

\author[0000-0002-2011-4924]{Genaro Su\'arez}
\affiliation{Department of Astrophysics, American Museum of Natural History, Central Park West at 79th St., New York, NY 10024, USA}
\email{gsuarez@amnh.org}

\author[0000-0003-4636-6676]{Eileen C. Gonzales}
\affiliation{Department of Physics and Astronomy, San Francisco State University, 1600 Holloway Avenue, San Francisco, CA 94132, USA}
\email{egonzales@sfsu.edu}

\author[0000-0001-6251-0573]{Jacqueline K. Faherty}
\affiliation{Department of Astrophysics, American Museum of Natural History, Central Park West at 79th St., New York, NY 10024, USA}
\email{jfaherty@amnh.org}

\author[0000-0003-0774-6502]{Jason J. Wang
\begin{CJK*}{UTF8}{gbsn}(王劲飞)\end{CJK*}}
\affil{Center for Interdisciplinary Exploration and Research in Astrophysics (CIERA), Northwestern University,
1800 Sherman Ave, Evanston, IL, 60201, USA}
\affil{Department of Physics and Astronomy, Northwestern University, 2145 Sheridan Rd, Evanston, IL 60208, USA}
\email{jason.wang@northwestern.edu}

\author[0000-0003-3050-8203]{Stanimir A. Metchev}
\affiliation{Department of Physics and Astronomy, The University of Western Ontario, 1151 Richmond St, London, ON N6A 3K7, Canada}
\email{smetchev@uwo.ca}

\author[0000-0001-7896-5791]{Dan Caselden}
\affiliation{Department of Astrophysics, American Museum of Natural History, Central Park West at 79th St., New York, NY 10024, USA}
\email{dancaselden@gmail.com}

\author[0000-0001-7780-3352]{Michael C. Cushing}
\affiliation{Ritter Astrophysical Research Center, Department of Physics \& Astronomy, University of Toledo, 2801 W. Bancroft St., Toledo, OH 43606, USA}
\email{michael.cushing@utoledo.edu}

\author[0000-0003-0398-639X]{Roman Gerasimov}
\affiliation{Department of Physics and Astronomy, University of Notre Dame, Nieuwland Science Hall, Notre Dame, 46556, Indiana, USA}
\email{rgerasim@nd.edu}

\begin{abstract}

Galactic archaeology relies on stellar kinematics and chemical abundances to identify various stellar populations and associations.
With JWST, Galactic archaeology of ancient metal-poor brown dwarfs is now possible through high-sensitivity, medium-resolution spectroscopy.
We present the first metal-poor brown dwarf radial velocity (RV) survey using JWST/NIRSpec G395H spectra ($\lambda/\Delta\lambda \sim$ 3,000) for a sample of {23} LTY dwarfs across a wide range of temperatures and metallicities.
We introduce a forward-modeling framework to measure robust RVs for brown dwarfs by focusing on the CO fundamental band at $\sim$4.5~{\micron}.
By comparing to sources with existing high-resolution RV measurements, we demonstrate that G395H RVs can achieve a systematic uncertainty of 5~{\kms},
sufficient to examine their Galactic space motions.
We identify kinematic members of the Milky Way's thin disk, thick disk, and halo populations in this sample.
We also confirm WISE J155349.98$+$693355.2 as likely associated with the Gaia-Enceladus merger remnant, 
and 2MASS J05325346$+$8246465 is likely associated with the Thamnos stream.
Our study systematically demonstrates that JWST G395H spectroscopy can provide robust RVs in addition to detailed chemical abundances, necessary for Galactic archaeology studies.
Future discoveries and characterization of brown dwarfs associated with Galactic halo substructures and streams will enable a stringent test of stellar/substellar formation and evolutionary models for diverse metallicities.

\end{abstract}


\section{Introduction} \label{sec:intro}

Galactic archaeology is the study of various stellar populations in the Milky Way, combining kinematic and chemical information.
Thanks to major surveys such as SDSS \citep{Abazajian:2009aa}, Gaia \citep{Gaia-Collaboration:2016aa}, GALAH \citep{De-Silva:2015aa}, APOGEE \citep{Majewski:2017aa}, H3 \citep{Conroy:2019aa}, DESI \citep{Cooper:2023aa}, 
major findings 
include the discoveries of in-situ and accreted halo components \citep{Nissen:2010aa}, the Gaia Enceladus dwarf galaxy merger remnant \citep{Belokurov:2018aa, Myeong:2018aa, Helmi:2018aa, Di-Matteo:2019aa, Gallart:2019aa}, dozens of stellar streams \citep{Helmi:2020aa, Bonaca:2025aa}, among others (see e.g., \citealp{Deason:2024aa} for a recent review).
These findings allow us to study the Milky Way's assembly history and mass distribution, as well as constrain dark matter distribution in the Milky Way's gravitational potential \citep{Helmi:2008aa, Ivezic:2012aa, Rix:2013aa, Belokurov:2013aa, Bland-Hawthorn:2016aa, Chakrabarti:2022aa}.

Galactic archaeology has largely focused on main sequence and giant stars as tracer populations. However, at the lowest masses, 
brown dwarfs provide an alternate population to study the evolution of the Milky Way system.
These objects are thought to form similarly to stars but have insufficient mass to sustain hydrogen fusion (M $\lesssim$ 0.075~{\msun}; \citealt{Kumar:1962aa, Kumar:1963aa, Hayashi:1963aa}).
Discovered just over three decades ago \citep{Nakajima:1995aa, Rebolo:1996aa},
we now know that brown dwarfs are ubiquitous in our Milky Way, but are more difficult to uncover and characterize than their brighter stellar counterparts \citep{Kirkpatrick:2005aa}. %
Studying substellar populations in the Milky Way through their kinematics and chemical composition enables a new means to study the Galaxy's formation and evolution history.
These substellar populations have essentially infinite lifetimes, and their relative abundances compared to lowest-mass stellar populations enable direct measurements of stellar/substellar mass function, birth rates, and chemical evolution \citep{Kirkpatrick:2005aa}.
These populations do not fuse species including hydrogen and lithium ($\lesssim$0.060~{\msun}; \citealp{Burrows:2001aa}) and are fully convective such that they preserve their initial chemical compositions, offering a unique and important probe into cosmic chemical evolution since the birth of the Milky Way.

Previous 3D kinematic surveys of brown dwarfs have largely focused on near-solar-metallicity, local thin disk populations \citep{Reid:2002aa, Zapatero-Osorio:2007aa, Blake:2010aa, Burgasser:2015ac, Hsu:2021aa, Hsu:2024aa}. The core limitation has been the need for high-resolution spectroscopy to measure radial velocities (RVs), restricting samples to %
relatively bright and nearby brown dwarfs (e.g., J or K $\lesssim$14.5--15.5~mag for Keck/NIRSPEC; \citealp{Hsu:2021aa}).
Nevertheless, these studies have provided important insights into the star formation and dynamical evolution of Galactic brown dwarfs.
As late-M and L dwarfs are a mixture of low-mass stars and brown dwarfs while late-L and T dwarfs are all brown dwarfs, their population velocity dispersions (cf. \citealp{Wielen:1977aa, Aumer:2009aa}) provide a clue to their ages and the Galactic star formation history at the very-low-mass end.
One key finding was the early indication of a surprisingly old population of local L dwarfs, inconsistent with population simulations \citep{Blake:2010aa, Burgasser:2015ac}.
A more complete kinematic study of late-M, L, and T dwarfs within 20~pc by \cite{Hsu:2021aa} resolved this discrepancy as being due to the contamination of unusually blue L dwarfs, an older and slightly metal-poor contaminant population \citep{Faherty:2009aa, Burgasser:2015ac}.
{That} study also revealed the first kinematic evidence of the stellar/substellar boundary around spectral type L4--L6, consistent with evolutionary models \citep{Baraffe:2003aa} and 
dynamical mass measurements \citep{Dupuy:2017aa}\footnote{See the discussions in their Section 6.3 of \cite{Hsu:2021aa} about the early determination of the stellar/substellar boundary at 2075~K, corresponding to L1--L2 by \cite{Dieterich:2014aa}.}.
Population simulations applied to these kinematic data also indicate a low-mass star formation history
consistent with exponential decline instead of a uniform star formation rate (see also, e.g., \citealp{Day-Jones:2013aa}).

While the local sample includes members of older, more dispersed populations from the Milky Way's thick disk and halo, they are present at low relative numbers (12\% and 0.5\%; \citealp{Juric:2008aa}). 
Metal-poor L subdwarfs at the stellar/substellar boundary were identified in early surveys \citep{Burgasser:2003ac, Burgasser:2004ac, Sivarani:2009aa, Zhang:2017aa, Zhang:2018aa} and shown to possess kinematics consistent with thick disk and halo populations \citep{Cushing:2009aa, Lodieu:2015ab}.  
More recently, several surveys have uncovered significantly metal-poor T and Y-type brown dwarfs in the local volume 
\citep{Zhang:2019aa, Schneider:2020aa, Meisner:2020aa, Meisner:2021aa, Kirkpatrick:2021ab, Schneider:2021aa, Burgasser:2025aa}, {substantially} expanding individual T subdwarf discoveries \citep{Burningham:2010aa, Pinfield:2012aa},  
whose intrinsically faint magnitudes have made it difficult to obtain accurate astrometric and RV measurements. 
In addition, deep surveys with HST 
\citep{Ryan:2005aa, Ryan:2011aa, Holwerda:2014aa, Ryan:2017aa, Aganze:2022aa}
and more recently JWST \citep{Nonino:2023aa, Burgasser:2024aa, Holwerda:2024aa, Hainline:2024ab, Chen:2025aa, Hainline:2026aa, Tu:2025aa, Morrissey:2026aa, Li:2026aa}
have begun to reach thick disk and halo brown dwarfs at kiloparsec distances.
The spectral morphologies of these low-temperature, metal-poor brown dwarfs provide a new stringent test of our understanding of cool atmospheres, brown dwarf evolution, and low-mass star formation \citep{Meisner:2021aa, Alvarado:2024aa, Gerasimov:2024ab, Mukherjee:2024aa}.

The nearest brown dwarfs, while containing few metal-poor populations, enable the detailed photometric and spectroscopic measurements needed for Galactic archaeology studies.  
Many of these sources have been validated with multi-epoch astrometry and proper motion surveys (e.g.,  \citealp{Marocco:2021aa, Meisner:2023ab, Schneider:2023ab, Marocco:2024aa, Karpov:2025aa, Schneider:2025aa, Zhang:2025ae, Marocco:2026aa}) 
yielding 5D kinematic information.
The last piece of the kinematics puzzle is radial velocity. 
While the faint magnitudes of these sources may limit our ability to obtain high-resolution spectroscopy, their highly structured spectra shaped by molecular absorption features {enable} RV precisions sufficient for kinematic studies ($\sigma_{RV} \sim $few~km/s) with lower-resolution spectrographs (R $\approx$ 3,000---6,000).

In this work, we present a brown dwarf RV survey based on observations 
of {23} L-, T-, and Y-type nearby brown dwarfs across a range of metallicities,
obtained with the JWST Near-Infrared Spectrograph (NIRSpec; \citealt{Jakobsen:2022aa}). 
Section~\ref{sec:data} describes our sample construction and observations, including
updated proper motions for two sources in our sample.
Section~\ref{sec:rv} illustrates our forward-modeling method to measure robust RVs, validated through comparison to high-resolution spectroscopic measurements in the literature.
Section~\ref{sec:kinematics} presents our kinematic analysis, including $UVW$ space motions, Galactic orbits, and population membership. We evaluate orbital energy and angular momentum to validate population assignments, as well as identify members of discrete Galactic substructures.
Notable individual objects are discussed in Section~\ref{sec:individual}.
We summarize our findings in Section~\ref{sec:sum}.

\section{Sample and Observations} \label{sec:data}

Our sample 
(Table~\ref{tab:sample}) is drawn from nearby low-mass stars and brown dwarfs with L, T, and Y type classifications observed with NIRSpec using its G395H high-resolution grating. 
We include both benchmark brown dwarf companions with solar or near-solar metallicities, and unassociated metal-poor brown dwarfs (subdwarfs `sd' and extreme subdwarfs `esd') in the field.
The majority of the targets are drawn from JWST General Observer (GO) program GO-4668 (PI: Burgasser); 
we also include 
the T8 dwarf 2MASSI J0415195-093506 (J0415$-$0935; \citealp{Burgasser:2002aa, Alejandro-Merchan:2025aa}) and the Y1 dwarf WISE J154151.65-225024.9 (J1541$-$2250; \citealp{Cushing:2011aa, Kirkpatrick:2011aa}) observed in GO-2124 (PI: Faherty), and
the Y0 dwarf WISE J182831.08+265037.7 (J1828$+$2650; \citealp{Cushing:2011aa, Kirkpatrick:2011aa, Lew:2024aa}) observed in program GO-1189 (PI: Roellig). 

JWST/NIRSpec observations are summarized in Table~\ref{tab:observations}.
All sources were observed using the NIRSpec's G395H grating and F290LP filter,
providing moderate-resolution spectra ($\lambda/\Delta\lambda$ $\approx$ 3,000) over 2.9--5.1~$\mu$m.
For program GO-4668, data were obtained using the S200A1 plus S200A2 slit combination, which 
covers the 3.7--3.8~$\mu$m gap between NIRSpec's two detectors.
Data for programs GO-1189 and GO-2124 were obtained using only the S200A1 slit.
We refer interested readers to \cite{Lew:2024aa}; \citet{Faherty:2025aa,Alejandro-Merchan:2025aa};  and \citet{Burgasser:2025ab}
for full descriptions of the observations.
All data were reduced using the JWST science calibration pipeline (versions 1.14.0, 1.15.1, 1.16.1, 1.17.1, 1.20.2) \citep{Bushouse:2026aa}, and downloaded from the Mikulski Archive for Space Telescopes (MAST) Portal\footnote{\url{https://mast.stsci.edu/}}.
The signal-to-noise ratio (SNR) per pixel in the spectral band of interest (4.4--5.0~{\micron}; see Section~\ref{sec:model}) ranges from 19 to 123, with a median of 56.
It is noted that the reduced JWST NIRSpec G395H spectra are already corrected for barycentric (up to $\pm$30~{\kms}) and spacecraft velocities (up to $\pm$1~{\kms}), with the corrections stored as the `VELOSYS' keyword in the reduced fits files for bookkeeping purposes (priv. comm. Cheryl Pavlovsky).

For completeness, we also analyzed the kinematics for the metal-poor Y dwarf WISEA J153429.75$-$104303.3 (J1534$-$1043; a.k.a. ``The Accident''; \citealp{Kirkpatrick:2021ab}) using the literature astrometry and radial velocity measurements from \cite{Faherty:2025aa}\footnote{The spectra of J1534$-$1043 show SiH$_4$ absorption in the 4.4--4.8~{\micron} \citep{Faherty:2025aa}, so the Sonora Elf-Owl models used in this work are unable to provide a good fit to J1534$-$1043.}.
Our findings of J1534$-$1043 are fully consistent with the halo kinematics indicated in \cite{Faherty:2025aa}, not associated with known dwarf galaxy merger remnants or streams.

\begin{deluxetable*}{lccccccccc}
\tablewidth{700pt}
\tablecaption{JWST NIRSpec/G359H Brown Dwarf Kinematics Sample \label{tab:sample}} 
\tabletypesize{\scriptsize} 
\tablehead{ 
\colhead{Name} & 
\colhead{Full Name} & 
\colhead{SpT} & 
\colhead{Cat.\tablenotemark{a}} & 
\colhead{RA} & 
\colhead{Dec} & 
\colhead{Parallax} & 
\colhead{$\mu_{\alpha}$} & 
\colhead{$\mu_{\delta}$} & 
\colhead{Ref.}
 \\
\colhead{} & \colhead{} & \colhead{} & \colhead{} & 
\colhead{(deg)} & \colhead{(deg)} & 
\colhead{(mas)} & \colhead{(mas yr$^{-1}$)} & \colhead{(mas yr$^{-1}$)} &\colhead{}
} 
\startdata
\hline
J0414$-$5854 & CWISEP J041451.75$-$585454.0 & esdT6 & sd & $63.716$ & $-58.915$ & $15 \pm 4$ & $212 \pm 11$ & $708 \pm 12$ & (3), (17), (9) \\ 
J0415$-$0935 & 2MASSI J0415195$-$093506 & T8 & d & $63.842$ & $-9.583$ & $175 \pm 2$ & $2214 \pm 1$ & $536 \pm 1$ & (2), (6) \\ 
J0448$-$1935 & WISE J044853.28$-$193548.6 & sdT5 & sd & $72.224$ & $-19.595$ & $58 \pm 3$ & $901.1 \pm 0.9$ & $761.1 \pm 0.9$ & (20), (12) \\ 
J0532$+$8246 & 2MASS J05325346$+$8246465 & esdL8 & sd & $83.303$ & $82.771$ & $40.7 \pm 0.5$ & $2038.8 \pm 0.6$ & $-1663.0 \pm 0.5$ & (3), (7) \\ 
J0645$-$6646 & 2MASS J06453153$-$6646120 & d/sdT0 & d/sd & $101.371$ & $-66.764$ & $54 \pm 3$ & $-885 \pm 2$ & $1312 \pm 2$ & (3), (12) \\ 
 HIP 38939 B & 2MASS J07580132$-$2538587 & d/sdT4.5 & d/sd & $119.507$ & $-25.651$ & $54.16 \pm 0.02$ & $362.41 \pm 0.01$ & $-245.79 \pm 0.02$ & (3), (7) \\ 
J0836$-$1859 & WISE J083641.10$-$185947.0 & d/sdT8 & d/sd & $129.171$ & $-18.997$ & $44 \pm 2$ & $-52.5 \pm 0.8$ & $-153.0 \pm 0.8$ & (20), (12) \\ 
J1316$+$0755 & ULAS J131610.28$+$075553.0 & sdT6.5 & sd & $199.042$ & $7.931$ & $17 \pm 3$ & $-1026 \pm 10$ & $107 \pm 10$ & (3), (5), (9) \\ 
J1416$+$1348 B & ULAS J141623.94$+$134836.3 & sdT7 & d/sd & $214.1$ & $13.81$ & $107.7 \pm 0.2$ & $86.7 \pm 0.3$ & $128.0 \pm 0.2$ & (3), (7) \\ 
J1416$+$1348 A & 2MASS J14162408$+$1348263 & d/sdL7 & d/sd & $214.101$ & $13.808$ & $107.7 \pm 0.2$ & $86.7 \pm 0.3$ & $128.0 \pm 0.2$ & (4), (7) \\ 
 HD 126053 B & ULAS J142320.79$+$011638.2 & sdT7.5 & sd & $215.837$ & $1.276$ & $57.27 \pm 0.04$ & $223.53 \pm 0.05$ & $-478.28 \pm 0.03$ & (3), (7) \\ 
 HIP 70849 B & 2MASS J14284235$-$4628393 & T4.5 & d & $217.176$ & $-46.478$ & $41.46 \pm 0.02$ & $-44.05 \pm 0.02$ & $-201.58 \pm 0.02$ & (13), (7) \\ 
 GJ 576 B & ULAS J150457.65$+$053800.8 & d/sdT5.5 & d/sd & $226.239$ & $5.632$ & $52.5 \pm 0.02$ & $-607.63 \pm 0.02$ & $-506.51 \pm 0.02$ & (3), (7) \\ 
 Gl 584 C & Gl 584 C & L8 & d & $230.845$ & $30.248$ & $56 \pm 0.8$ & $116.8 \pm 0.4$ & $-171.4 \pm 0.5$ & (10), (18) \\ 
J1541$-$2250 & WISE J154151.65$-$225024.9 & Y1 & d & $235.464$ & $-22.841$ & $167 \pm 2$ & $-902.8 \pm 0.9$ & $-91.4 \pm 0.9$ & (11), (12) \\ 
J1553$+$6933 & WISE J155349.98$+$693355.2 & sdT4 & sd & $238.459$ & $69.565$ & $26 \pm 3$ & $-1527 \pm 12$ & $1258 \pm 13$ & (3), (16), (21) \\ 
J1626$+$3925 & 2MASS J16262034$+$3925190 & usdL4 & sd & $246.576$ & $39.423$ & $32.3 \pm 0.2$ & $-1374.8 \pm 0.2$ & $237.4 \pm 0.2$ & (19), (7) \\ 
J1810$-$1010 & CWISEP J181006.00$-$101001.1 & esdT3 & sd & $272.526$ & $-10.167$ & $112 \pm 8$ & $-1027 \pm 4$ & $-246 \pm 4$ & (3), (14), (22) \\ 
J1828$+$2650 & WISE J182831.08$+$265037.7 & Y0 & d & $277.13$ & $26.844$ & $100 \pm 2$ & $1016.5 \pm 0.8$ & $169.3 \pm 0.8$ & (11), (8) \\ 
 Wolf 1130 C & Wolf 1130 C & sdT8 & sd & $301.335$ & $54.409$ & $60.3 \pm 0.03$ & $-1159.52 \pm 0.04$ & $-904.01 \pm 0.03$ & (3), (7) \\ 
J2030$+$0749 & 2MASS J20304235$+$0749358 & T1.5 & d & $307.68$ & $7.826$ & $102.8 \pm 0.8$ & $664.0 \pm 0.9$ & $-111.7 \pm 0.9$ & (1), (7) \\ 
 HN Peg B & HN Peg B & T2.5 & d & $326.12$ & $14.768$ & $55.15 \pm 0.03$ & $231.11 \pm 0.03$ & $-113.2 \pm 0.03$ & (15), (7) \\ 
\hline
\multicolumn{10}{c}{Additional Literature Metal-poor Brown Dwarf} \\
\hline
 J1534$-$1043 & WISEA J153429.75$-$104303.3 & esdY & sd & $233.622$ & $-10.722$ & $61 \pm 5$ & $-1253 \pm 9$ & $-2377 \pm 7$ & (23), (24) \\ 
\enddata
\tablerefs{(1) \cite{Best:2013aa}; (2) \cite{Burgasser:2002aa}; (3) \cite{Burgasser:2025aa}; (3) \cite{Burgasser:2025aa}; (3) \cite{Burgasser:2025aa}; (4) \cite{Burningham:2010aa}; (5) \cite{Burningham:2014aa}; (6) \cite{Dupuy:2012aa}; (7) \cite{Gaia-Collaboration:2023aa}; (8) \cite{Hsu:2021aa}; (9) This work; (10) \cite{Kirkpatrick:2000aa}; (11) \cite{Kirkpatrick:2011aa}; (12) \cite{Kirkpatrick:2021aa}; (13) \cite{Lodieu:2014aa}; (14) \cite{Lodieu:2022aa}; (15) \cite{Luhman:2007aa}; (16) \cite{Meisner:2020aa}; (17) \cite{Schneider:2020aa}; (18) \cite{van-Leeuwen:2007aa}; (19) \cite{Zhang:2013aa}; (20) \cite{Zhang:2019aa}; (21) \cite{Zhang:2025ac}; (22) \cite{Zhang:2025ad}; (23) \cite{Kirkpatrick:2021ab}; (24) \cite{Faherty:2025aa}.}
\tablenotetext{a}{We organize our sample into three categories: 
benchmark solar-metallicity brown dwarfs (d), 
metal-poor brown dwarfs (sd), 
and mild subdwarfs (d/sd).}
\end{deluxetable*}

\begin{deluxetable*}{lcccccccccc}
\tablewidth{700pt}
\tablecaption{NIRSpec Observations \label{tab:observations}} 
\tabletypesize{\scriptsize} 
\tablehead{ 
\colhead{Name} & 
\colhead{Desination} & 
\colhead{$W$1} & 
\colhead{$W$2} & 
\colhead{GO} & 
\colhead{ID} & 
\colhead{Date of Obs.} & 
\colhead{Slit \tablenotemark{a}} & 
\colhead{Exposure} & 
\colhead{Reduction} & 
\colhead{SNR \tablenotemark{b}}
 \\
\colhead{} & \colhead{(J2000)} & \colhead{(mag)} & \colhead{(mag)} & \colhead{ID} & \colhead{ID} & \colhead{(UT)} & \colhead{} & \colhead{(s)} & 
\colhead{Version} & \colhead{}
} 
\startdata
\hline
J0414$-$5854 & J041451.80$-$585452.4 & $16.76 \pm 0.03$ & $15.36 \pm 0.03$ & 4668 & 72 & 2025-01-14 & S200A1+S200A2 & 1538 & 1.16.1 & 48 \\ 
J0415$-$0935 & J041522.17$-$093457.1 & $15.15 \pm 0.03$ & $12.3 \pm 0.01$ & 2124 & 59 & 2022-10-16 & S200A1 & 154 & 1.14.0 & 35 \\ 
J0448$-$1935 & J044853.70$-$193543.6 & $16.3 \pm 0.03$ & $14.23 \pm 0.02$ & 4668 & 103 & 2024-08-30 & S200A1+S200A2 & 879 & 1.15.1 & 57 \\ 
J0532$+$8246 & J053312.61$+$824617.2 & $13.85 \pm 0.01$ & $13.27 \pm 0.01$ & 4668 & 19 & 2024-09-23 & S200A1+S200A2 & 659 & 1.15.1 & 79 \\ 
J0645$-$6646 & J064529.10$-$664550.7 & $13.74 \pm 0.02$ & $13.3 \pm 0.02$ & 4668 & 10 & 2024-08-25 & S200A1+S200A2 & 659 & 1.14.0 & 73 \\ 
 HIP 38939 B & J075801.75$-$253903.1 & $15.77 \pm 0.04$ & $13.86 \pm 0.02$ & 4668 & 101 & 2024-11-24 & S200A1+S200A2 & 659 & 1.15.1 & 54 \\ 
J0836$-$1859 & J083641.13$-$185947.7 & $17.63 \pm 0.09$ & $15.13 \pm 0.03$ & 4668 & 106 & 2024-11-24 & S200A1+S200A2 & 1319 & 1.15.1 & 19 \\ 
J1316$+$0755 & J131610.03$+$075551.5 & $16.54 \pm 0.04$ & $15.95 \pm 0.06$ & 4668 & 109 & 2025-02-19 & S200A1+S200A2 & 1758 & 1.16.1 & 39 \\ 
J1416$+$1348 B & J141623.96$+$134837.3 & $16.05 \pm 0.03$ & $12.83 \pm 0.01$ & 4668 & 43 & 2025-02-28 & S200A1+S200A2 & 440 & 1.16.1 & 57 \\ 
J1416$+$1348 A & J141624.16$+$134828.1 & $11.35 \pm 0.02$ & $11.02 \pm 0.01$ & 4668 & 6 & 2025-02-26 & S200A1+S200A2 & 176 & 1.16.1 & 93 \\ 
 HD 126053 B & J142320.90$+$011635.3 & $17.9 \pm 0.1$ & $14.78 \pm 0.02$ & 4668 & 112 & 2025-02-28 & S200A1+S200A2 & 1099 & 1.17.1 & 48 \\ 
 HIP 70849 B & J142842.28$-$462842.5 & $15.17 \pm 0.02$ & $13.84 \pm 0.02$ & 4668 & 85 & 2025-02-28 & S200A1+S200A2 & 659 & 1.16.1 & 47 \\ 
 GJ 576 B & J150457.32$+$053756.8 & $16.21 \pm 0.03$ & $14.31 \pm 0.02$ & 4668 & 40 & 2025-02-28 & S200A1+S200A2 & 879 & 1.16.1 & 53 \\ 
 Gl 584 C & J152322.78$+$301453.4 & $13.5 \pm 0.01$ & $13.02 \pm 0.01$ & 4668 & 80 & 2025-02-27 & S200A1+S200A2 & 440 & 1.16.1 & 60 \\ 
J1541$-$2250 & J154151.41$-$225026.1 & $17.6 \pm 0.2$ & $14.22 \pm 0.03$ & 2124 & 5 & 2023-03-05 & S200A1 & 673 & 1.14.0 & 34 \\ 
J1553$+$6933 & J155350.23$+$693355.7 & $17.07 \pm 0.03$ & $15.62 \pm 0.03$ & 4668 & 32 & 2024-11-21 & S200A1+S200A2 & 1319 & 1.20.2 & 35 \\ 
J1626$+$3925 & J162618.23$+$392523.4 & $13.496 \pm 0.009$ & $13.165 \pm 0.008$ & 4668 & 26 & 2024-08-19 & S200A1+S200A2 & 440 & 1.14.0 & 65 \\ 
J1810$-$1010 & J181006.12$-$101001.5 & $13.65 \pm 0.02$ & $12.483 \pm 0.009$ & 4668 & 35 & 2024-08-19 & S200A1+S200A2 & 659 & 1.14.0 & 123 \\ 
J1828$+$2650 & J182831.08$+$265037.7 & $18.8 \pm 0.2$ & $14.39 \pm 0.02$ & 1189 & 11 & 2022-07-28 & S200A1 & 2276 & 1.14.0 & 97 \\ 
 Wolf 1130 C & J200520.38$+$542433.9 & $17.16 \pm 0.04$ & $15.14 \pm 0.02$ & 4668 & 52 & 2024-08-30 & S200A1+S200A2 & 1099 & 1.15.1 & 57 \\ 
J2030$+$0749 & J203043.10$+$074933.9 & $13.02 \pm 0.02$ & $12.2 \pm 0.02$ & 4668 & 29 & 2025-05-01 & S200A1+S200A2 & 440 & 1.17.1 & 93 \\ 
 HN Peg B & J214428.75$+$144606.4 & $13.2 \pm 0.03$ & $12.6 \pm 0.02$ & 4668 & 83 & 2024-10-16 & S200A1+S200A2 & 440 & 1.15.1 & 44 \\ 
\enddata
\tablerefs{{$W$1 and $W2$: J1810$-$1010: \cite{Schneider:2020aa}; Wolf 1130C: \cite{Zhang:2025ac}; HN Peg B: AllWISE \citep{Cutri:2014aa}; the rest of the sample: CatWISE \citep{Eisenhardt:2020aa} and CatWISE2020 \citep{Marocco:2021aa}.}}
\tablenotetext{a}{All data are observed using the G395H/F290LP grating/filter setting.}
\tablenotetext{b}{The median signal-to-noise ratio (SNR) per pixel in the range 4.4--5.0~{\micron}.}
\end{deluxetable*}

\subsection{New Proper Motion Measurements} \label{sec:pm}

For the majority of our sample, we used the highest precision measurements from the literature for parallax and proper motions (Table~\ref{tab:sample}). 
For two sources, WISE J0414$-$5854 and ULAS J1316$+$0755, we report new and improved proper motion measurements based on combined astrometry from multiple surveys.

We then conducted a linear fit to all epochs to derive the proper motions, following the method described in \cite{Schneider:2023ab, Schneider:2026aa}.
For J0414$-$5854, we combined seven $z$-band and $y$-band detections from NOIRLab Source Catalog DR2 \citep{Nidever:2021ab}, three $J$-band detections from VISTA Hemisphere Survey (VHS) \citep{McMahon:2013aa}, and 32 $W1$- and $W2$-band detections from Wide-field
Infrared Survey Explorer (WISE)/unTimely \citep{Meisner:2023ab}, spanning a total of 10.5 years.
Linear fits to the astrometry yield proper motions of $\mu_{\alpha} = 212 \pm 11$~mas yr$^{-1}$ and $\mu_{\delta} = 708 \pm 12$~mas yr$^{-1}$, compared to $\mu_{\alpha} = 223 \pm 26$~mas yr$^{-1}$ and $\mu_{\delta} = 672 \pm 24$~mas yr$^{-1}$ reported in \cite{Marocco:2021aa}.
For J1316$+$0755, early WISE/unTimely detections are blended with a background source, so we used only epochs after 2014 (13 detections). We combined with eight UKIRT Infrared Deep Sky Survey (UKIDSS) $Y$- and $J$-band detections \citep{Lawrence:2007aa}, spanning a total of 14 years from 2006 (VHS) to 2020 (WISE).
These data yield $\mu_{\alpha} = -1026 \pm 10$~mas yr$^{-1}$ and $ \mu_{\delta} = 107 \pm 10$~mas yr$^{-1}$, compared to $ \mu_{\alpha} = - 1012 \pm 15 $~mas yr$^{-1}$ and $ \mu_{\delta} = 103 \pm 14$~mas yr$^{-1}$ in \cite{Burningham:2014aa}.
Our new proper motion measurements are fully consistent with literature values and show 1.4 to 2.4 times improvements in precision.
{We note that the quality of these imaging} data {was} not sufficient to provide solid constraints on the trigonometric parallaxes {of these two sources}.

\section{Radial Velocity Measurements} \label{sec:rv}

\subsection{Forward-modeling Method} \label{sec:model}

\begin{figure*}
    \centering
    \includegraphics[trim=0 0.55cm 0 0.62cm, width=0.90\textwidth]{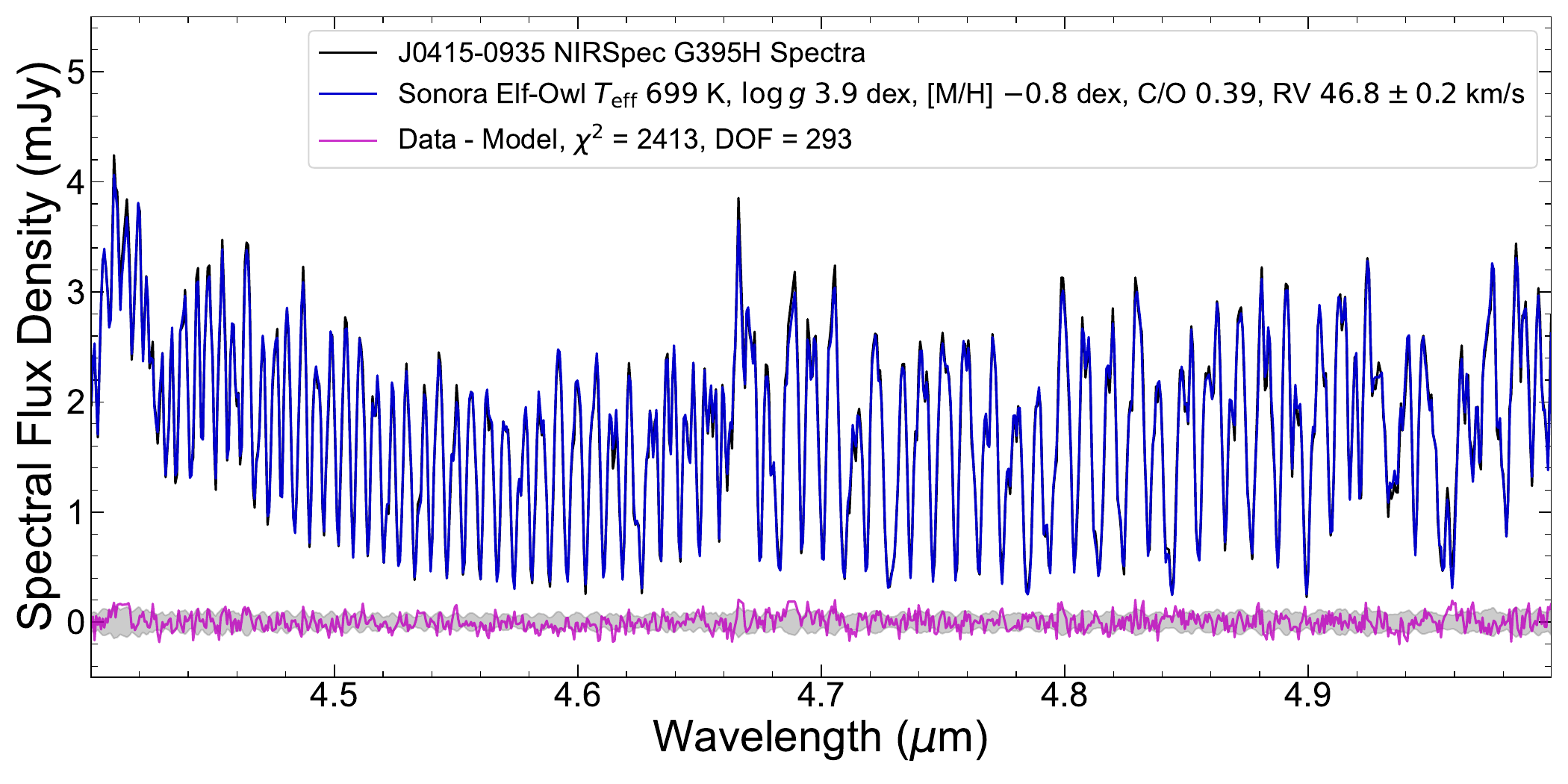}
    \caption{Example model fit to the JWST/NIRSpec G395H spectrum of the T8 dwarf J0415$-$0935 , with relatively low SNR $\sim$35 compared to our full sample (Table~\ref{tab:observations}). 
    The observed spectrum is shown in black, with the corresponding $\pm$1$\sigma$ uncertainty indicated by the grey shaded region centered at zero.
    The best-fit Sonora Elf-Owl model \citep{Mukherjee:2024aa} is shown in blue, and
    the difference spectrum (data $-$ model) is shown in magenta.
    The majority of structure observed in this region arises from the CO (1$-$0) fundamental band, which is used to anchor our RV measurements.
    Fits for each source in our full sample {(22 objects)} are provided as a figure set in the online journal.
    }\label{fig:example_fit}
\end{figure*}

The NIRSpec G395H data allow us to derive the full 6D phase space coordinate of our sources by providing the necessary radial velocity.
We deployed a forward-modeling approach based on the established method of modeling
the highly-structured CO first overtone (2$-$0) rotational-vibrational band at $\sim$2.3~{\micron} in late-M, L, and T dwarf near-infrared spectra 
\citep[e.g.,][]{Blake:2007aa,Blake:2010aa, Burgasser:2015ab, Burgasser:2016aa, Hsu:2021aa, Hsu:2023aa}.
Fortunately, NIRSpec G395H spectra encompass the similarly structured CO fundamental band from 4.4 to 5.0~{\micron}
(Figure~\ref{fig:example_fit}).
Our analysis is based on the method and the \texttt{SMART} code described in \citet{Hsu:2021aa,Hsu:2023aa}, excluding fit to telluric absorption.
Briefly, 
we used the Sonora Elf-Owl atmosphere models \citep{Mukherjee:2024aa} ($\lambda/\Delta\lambda$ $\sim$ 60,000), allowing for linear interpolation over effective temperature ({\teff}), surface gravity ({\logg}), metallicity ([M/H]), C/O ratio, and vertical eddy diffusion coefficient ($\kappa_\mathrm{zz}$). 
The baseline model is modified by applying an RV shift and rotational broadening, the latter using the
fast implementation kernel described in \cite{Carvalho:2023aa}. The model is then convolved with a Gaussian instrumental line-spread function using the \texttt{PyAstronomy} ``instrBroadGaussFast'' function \citep{Czesla:2019aa}, and resampled onto the wavelength grid of the data in the 4.4--5.0~{\micron} band while preserving the total flux (using the \texttt{SMART} ``integralResample'' function).
Additionally, we include an overall flux scale factor ($\sigma_\mathrm{flux}$) for the model, and a noise jitter term ($\sigma_\mathrm{noise}$) as a fraction of the observed flux added to the data error in quadrature.
Model and observed spectra are compared using a chi-square statistic, with a 2$^\mathrm{nd}$-order polynomial continuum correction applied to minimize this statistic.
We used the Markov Chain Monte Carlo (MCMC) sampler \texttt{emcee} \citep{Foreman-Mackey:2013aa} 
to explore the parameter space and obtain best-fit values and uncertainties for the parameters and priors listed in Table~\ref{table:mcmc_parameter}.
For this analysis, we deployed 100 MCMC walkers and 2000 total steps per walker, with the first 1000 steps discarded as burn-in.

\begin{deluxetable}{lccc}
\tablecaption{Forward Modeling Parameters \label{table:mcmc_parameter}}
\tablecolumns{4}
\tablehead{
\colhead{Description} &  \colhead{Symbol (unit)}  & \colhead{Priors\tablenotemark{a}} & \colhead{Bounds}
}
\startdata
Temperature & $T_{\mathrm{eff}}$ (K) & (300, 1400) & (300, 2400) \\
Surface gravity & $\log{g}$ (cm s$^{-2}$) & (3.23, 5.5) & (3.23, 5.5) \\
Metallicity & [M/H] (dex) & ($-1.0$, $+1.0$) & ($-1.0$, $+1.0$) \\
C/O & C/O\tablenotemark{b} & (0.229, 1.145) & (0.229, 1.145) \\
Vertical diffusion & $\kappa_{zz}$ (cm$^2$ s$^{-1}$) & (10$^2$, 10$^8$) & (10$^2$, 10$^8$) \\
Rotational velocity & {\vsini} ({\kms}) & (0, 200) & (0, 200) \\
Radial velocity & RV ({\kms}) & ($-$250, $+$250) & ($-$400, $+$400) \\
Resolution & $R$ & (1500, 4000) & (1500, 4000) \\
Flux Scaling & $\sigma_\mathrm{flux}$ & (0.0, 1.0) & (0.0, 1.0) \\
Noise Jitter & $\sigma_\mathrm{noise}$ & (0.0, 0.5) & (0.0, 0.5)
\enddata
\tablenotetext{a}{All priors assume a uniform distribution over the range specified}
\tablenotetext{b}{The solar elemental abundance of C/O = 0.458, so our priors correspond to 0.5 to 2.5 $\times$ the solar C/O value \citep{Lodders:2009aa, Mukherjee:2024aa}.}
\end{deluxetable}

Figure~\ref{fig:example_fit} illustrates an example fit for the T8 dwarf J0415$-$0935.
The best-fit physical parameters are labeled in the legend, with sampling uncertainties without systematics across various models.
Our {\teff} = 699$^{+2}_{-4}$~K is fully consistent with the {\teff} = 677 $\pm$ 56~K from \cite{Filippazzo:2015aa}. However, our {\logg} = 3.9~cgs dex {differs significantly from} the {\logg} = 4.8 $\pm$ 0.5~cgs dex in \cite{Filippazzo:2015aa}, and metallicity [M/H] = $-$0.8~dex is {unphysical given that J0415$-$0935 is} a solar-metallicity T dwarf standard \citep{Burgasser:2006aa}.
Indeed, fitting the relatively narrow wavelength range generally exhibits systematic biases across several studies in the literature, especially high-resolution spectroscopic studies focused on optimizing the radial velocity measurements (see \citealp{Del-Burgo:2009aa, Hsu:2021aa, Hsu:2024aa, Hsu:2024ab}).
However, the main goal of this study is to obtain a reliable and robust radial velocity measurement, and the core of RV measurements is to obtain the best spectral template that matches the observed spectra.
While there are slight mismatches at 4.40--4.42~{\micron} and 4.66--4.70~{\micron}, our best-fit {model} show a reasonable fit to the observed spectra, with a reduced $\chi^2$ $\sim$ 8.07 (without the error jitter term).
From this analysis, we obtained a best-fit RV = $46.8\pm0.2$~{\kms}, where the uncertainty reflects the statistical distribution from our Bayesian posterior.
{To explore whether different {\teff}, {\logg}, and metallicity would bias our resulting RV, we conducted a restricted MCMC forward-modeling routine for J0415$-$0935, under the same method but fixing {\teff} = 677~K, {\logg} = 4.83~cgs dex, and solar metallicity from \cite{Filippazzo:2015aa}.
This experiment measured an RV = $46.8^{+0.4}_{-0.3}$~{\kms}, fully consistent with our reported best-fit RV ($46.8\pm0.2$~{\kms}), though with $\sim$1.5$\times$ larger uncertainty and a significantly poorer fit ($\chi^2 \approx 6335$).
This demonstrates that the CO and H$_2$O lines provide a robust RV anchor even if their atmospheric parameters might not be fully accurate.}

For most of our sample, statistical errors are below 0.5~{\kms}; the two exceptions are the extremely metal-poor subdwarfs usdL4 J1626+3925 and esdT3 J1810$-$1010 which have weak molecular absorption features and higher statistical errors of 1~{\kms}.
As discussed below, we include an additional 5~{\kms} systematic uncertainty to the measurement to account for calibration effects, which is 20 times smaller than the native resolution of the spectrum (or 1/10$^\mathrm{th}$ of the pixel). 
The combined value, 47$\pm$5~{\kms}, is fully consistent with the best ground-based measurement of RV = $51.9\pm1.1$~{\kms} reported by \cite{Hsu:2021aa} based on high-resolution ($\lambda/\Delta\lambda \approx 25,000$) Keck/NIRSPEC observations;
and with the RV = $47.1\pm1.8$~{\kms} reported by \cite{Alejandro-Merchan:2025aa} using the same JWST NIRSpec G395H spectra (which they used different wavelength segments and three substellar atmosphere models).

\subsection{New High-resolution Spectroscopic Radial Velocity Measurements}\label{sec:hrs_rv}

We also present new RV measurements for J1416$+$1348A using Keck/NIRSPEC high-resolution spectra \citep{McLean:1998aa, McLean:2000aa}.
Our NIRSPEC data for J1416$+$1348A were observed on 2015 December 29 (UT) (PI: Burgasser) under excellent conditions, with clear sky and excellent seeing of 0$\farcs$44.
The N7 filter (corresponding to the $K$-band) and the 0$\farcs$432$\times$12 slit were selected.
The spectra were taken under two 500-s exposures in an AB nodding pair.
We also observed the A0V HD 35505 to serve as our wavelength calibrator.
The data were reduced using the modified \texttt{NSDRP} code \citep{Tran:2016aa} detailed in \cite{Hsu:2021aa, Theissen:2022aa}.

The reduced spectra were then forward-modeled following the prescription of \cite{Hsu:2021aa, Hsu:2023aa} using \texttt{SMART} \citep{Hsu:2021ab} and the Markov Chain Monte Carlo \texttt{emcee} sampler \citep{Foreman-Mackey:2013aa}.
We chose the BT-Settl atmosphere models \citep{Allard:2012ab} and ESO Earth atmosphere models \citep{Moehler:2014aa}.
Our best-fit parameters for J1416$+$1348A are RV = $-$42.3$\pm$0.4~{\kms}, {\teff} = 1731$^{+21}_{-20}$~K, {\logg} = 5.50$^{+0.02}_{-0.04}$~cgs dex, and {\vsini} = 40.2$\pm$0.6~{\kms}.

\subsection{Validation of Radial Velocity Measurements}\label{sec:check}

Our RV measurements range from $-260$~{\kms} to $+130$~{\kms}, larger than the typical range ($|\mathrm{RV}| \leq$80~{\kms}) for nearby thin disk low-mass stars and brown dwarfs \citep{Blake:2010aa,Hsu:2021aa,Hsu:2024aa}.
To validate these RVs, we compared them to available literature measurements summarized in Table~\ref{tab:rv_uvw}.
In general, high-resolution ($\lambda/\Delta\lambda > 20,000$) spectroscopic RVs are the most reliable measurements, as the molecular lines are well-resolved and calibration is typically reliable to 0.1--0.5~{\kms} \citep{Hsu:2021aa, Hsu:2024aa}.
However, given the faintness of our sources, we also included medium-resolution ($3,000 \lesssim \lambda/\Delta\lambda \lesssim 6,000$) 
measurements where available.

\begin{figure}[ht!]
    \centering
    \includegraphics[trim=10 0cm 10 0cm, width=0.45\textwidth]{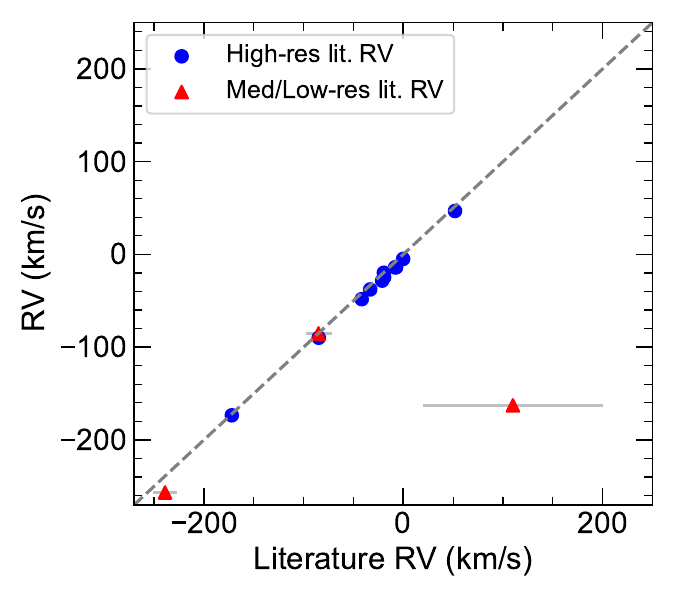}
    \caption{Comparison of our NIRSpec RV measurements with high-resolution ($\lambda/\Delta\lambda >$20,000; blue circles) 
    and moderate-resolution (3,000 $<\lambda/\Delta\lambda <$10,000; red triangles) spectroscopic measurements from the literature.
    The dashed line indicates perfect agreement.
    The source with the largest disagreement is J1553+6933, which has a very low precision measurement ($-$110$\pm$90~{\kms}) based on moderate-resolution spectra \citep{Burgasser:2025aa}.
    Our RV measurements show an excellent agreement with the literature high-resolution RVs (standard deviation of 2.3~{\kms} and median difference of $-$5.1~{\kms}, demonstrating that our method is robust using G395H at 4.4--5.0~{\micron} at $R \sim $3,000 resolution.
    }\label{fig:rv_literature_compare}
\end{figure}

Figure~\ref{fig:rv_literature_compare} illustrates this comparison.
We find overall agreement between our NIRSpec RVs and high-resolution RVs from the literature,
with a median difference (literature RV $-$ NIRSpec RV) of $+$5.1$\pm$2.1~{\kms}.
The source with the largest RV disagreement, the sdT4 J1553+6933, has a very low-precision measurement ($-$110$\pm$90~{\kms}) based on moderate-resolution spectra \citep{Burgasser:2025aa}, and we discount that measurement from our analysis\footnote{{The currently known (non-contacting) shortest orbital period ultracool dwarf binary LP 413-53 has a maximum RV difference of $\sim$52~{\kms}, with a period of 17~hr and separation $a < 20$ stellar radii \citep{Hsu:2023aa}.}}.
In particular, the medium-resolution $R \sim 2,700$ Keck/NIRES spectra of J1553+6933 from \citep{Burgasser:2025aa} have lower signal-to-noise ratios and telluric absorption from the ground, and the Keck/NIRES wavelength calibration is less reliable than the JWST NIRSpec G395H wavelength used in this study.
The marginal systematic offset, with individual values ranging between +0.3~{\kms} and +7.1~{\kms} (i.e., our JWST NIRSpec RVs are systematically more negative compared to the literature high-resolution RVs),
may be due to the pipeline wavelength calibration of NIRSpec data.
Our finding here is consistent with the JWST validation report on individual emission lines from unresolved, high-velocity planetary nebulae, and indicates an uncertainty of $\sim$6~{\kms} for G395H spectroscopy \citep{2025jwst.rept.9239G}.
Indeed, subpixel centering offsets of sources with the NIRSpec slit can result in velocity shifts of the same scale; a 0.1 pixel offset corresponds to a shift of 
approximately 5~{\kms}.
Thus, we adopt a systematic RV uncertainty of 5~{\kms} to our measurements for the remainder of the analysis.

\begin{figure}
    \centering
    \includegraphics[width=\linewidth]{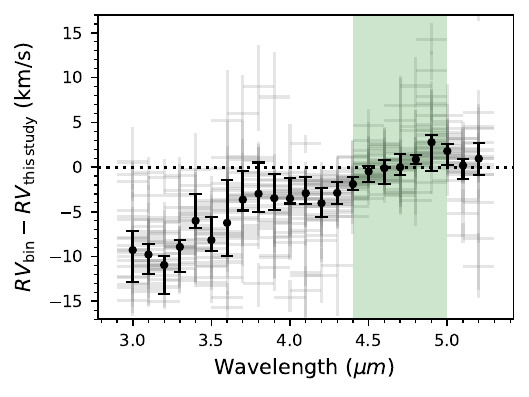}
    \caption{
    {Difference between RVs measured at different wavelength {bins ($RV_\mathrm{bin}$; 0.6\,{\micron} bin size in steps of 0.1\,{\micron}}) and the RV adopted in this study based on the CO band ($RV_\mathrm{this \, study}$; green shaded region, 4.4--5.0\,{\micron}), across the entire NIRSpec G395H wavelength range}. 
    Gray points show individual RV measurements for each source in each 0.6~$\mu$m bin, whereas
    black points with error bars indicate the median and 25$^\mathrm{th}$ to 75$^\mathrm{th}$ percentiles for each bin
    across all sources. 
    There is a clear shift toward more negative RVs at shorter wavelengths, but around the CO-band region used for our analysis RV values are roughly constant with wavelength.
    }\label{fig:rv_compaison}
\end{figure}

We also explored systematic errors driven by the choice of wavelength region used to measure our RVs. 
For every spectrum, we computed RVs using our forward modeling method using bins of 0.6~{\micron} in steps of 0.1~{\micron}, similar to the wavelength bin size used in our fitting routine, across the full 2.9--5.7~{\micron} range of the G395H data.
Results of this analysis are shown in Figure~\ref{fig:rv_compaison}.
There is a consistent offset and larger uncertainties for RVs derived at bluer wavelengths ($\lambda < 3.8$~{\micron}), 
but in the region of the CO band these differences level off at zero.
The larger uncertainties at wavelengths outside the CO band can be attributed to the absence of 
strong, dense, and well-structured molecular opacity features, particularly near the 4~{\micron} flux peak, 
whereas errors and offsets at shorter wavelengths may arise from saturated CH$_4$ absorption centered at 3.3~{\micron} (which generally show lower SNR than CO band at 4.5~{\micron}) for the T and Y dwarfs in the sample (cf.~\citealt{Faherty:2024aa, Lew:2024aa}), and/or wavelength-dependent systematics from the JWST wavelength calibration model.

{While RV measurements of young cool stars are impacted by stellar activity ($\sim$1--2~{\kms}, e.g., \citealp{Gully-Santiago:2017aa, Tang:2024aa}), our sample is largely T and Y dwarfs at field ages and much cooler temperatures ($\lesssim$1200~K).
Although brown dwarf variability is commonly found in young L-type brown dwarfs \citep{Vos:2022aa}, older and cool T/Y dwarfs show much less RV variability from high-resolution spectra \citep{Hsu:2021aa}.
A well-known young early T dwarf, J0136$+$0933 RV variations are $\lesssim$2~{\kms} \citep{Hsu:2021aa}, so our old brown dwarfs are expected to have RV variations less than this level (unless they are in a close binary; \citealp{Hsu:2023aa, Xuan:2024ac, Whitebook:2026aa}), and our systematic RV uncertainty of 5~{\kms} is larger than these RV variations.
}

In summary, we find good agreement between RVs measured for L, T and Y dwarfs between JWST/NIRSpec G395H observations and high-resolution spectroscopic observations to within a 5~{\kms} systematic uncertainty,
corresponding to 1/20$^\mathrm{th}$ of the resolution of the data. 

\begin{deluxetable*}{lccccccc}
\tablewidth{700pt}
\tablecaption{Measured Radial Velocities and Derived UVW Velocities \label{tab:rv_uvw}} 
\tabletypesize{\scriptsize} 
\tablehead{ 
\colhead{Name} & 
\colhead{RV\tablenotemark{a}} & 
\colhead{RV$_\mathrm{lit}$} & 
\colhead{RV$_\mathrm{lit}$} & 
\colhead{$U$} & 
\colhead{$V$} & 
\colhead{$W$} & 
\colhead{Pop.\tablenotemark{c}} \\
\colhead{} & \colhead{({\kms})} & \colhead{({\kms})} & 
\colhead{Ref.} & \colhead{({\kms})} & 
\colhead{({\kms})} & \colhead{({\kms})} & \colhead{}
} 
\startdata
\hline
J0414$-$5854 & $130.3^{+0.4}_{-0.5} \pm 5$ & \nodata & \nodata & $-224 \pm 68$ & $-76 \pm 5$ & $-89 \pm 5$ & TD/Halo \\ 
J0415$-$0935 & $46.8 \pm 0.2 \pm 5$ & $52 \pm 1$ & (4) & $-51 \pm 4$ & $-30 \pm 2$ & $24 \pm 3$ & D \\ 
J0448$-$1935 & $92.0 \pm 0.3 \pm 5$ & \nodata & \nodata & $-109 \pm 5$ & $-41 \pm 3$ & $28 \pm 5$ & D/TD \\ 
J0532$+$8246 & $-173.4 \pm 0.2 \pm 5$ & $-172 \pm 1$ & (8) & $-61 \pm 4$ & $-329 \pm 4$ & $57 \pm 3$ & Halo \\ 
J0645$-$6646 & $-25.4 \pm 0.3 \pm 5$ & $-33 \pm 10$\tablenotemark{b} & (10) & $-122 \pm 7$ & $43 \pm 5$ & $-32 \pm 3$ & TD \\ 
 HIP 38939 B & $-14.0 \pm 0.2 \pm 5$ & $-8.2 \pm 0.1$ & (2) & $49 \pm 2$ & $9 \pm 5$ & $22.5 \pm 0.2$ & D \\ 
J0836$-$1859 & $21.6^{+0.5}_{-0.6} \pm 5$ & \nodata & \nodata & $9 \pm 2$ & $-14 \pm 4$ & $-2 \pm 1$ & D \\ 
J1316$+$0755 & $24.8 \pm 0.4 \pm 5$ & \nodata & \nodata & $-229 \pm 44$ & $-137 \pm 25$ & $68 \pm 8$ & TD/Halo \\ 
J1416$+$1348 B & $-51.5 \pm 0.2 \pm 5$ & \nodata & \nodata & $-10 \pm 2$ & $17.8 \pm 0.1$ & $-40 \pm 5$ & D \\ 
J1416$+$1348 A & $-48.2 \pm 0.2 \pm 5$ & $-42.3 \pm 0.4$ & (5) & $-9 \pm 2$ & $17.9 \pm 0.1$ & $-37 \pm 5$ & D \\ 
 HD 126053 B & $-24.8 \pm 0.3 \pm 5$ & $-19.21 \pm 0.03$ & (9) & $30 \pm 3$ & $-2.3 \pm 0.6$ & $-37 \pm 4$ & D \\ 
 HIP 70849 B & $-4.9 \pm 0.2 \pm 5$ & $-0.13 \pm 0.03$ & (9) & $2 \pm 4$ & $2 \pm 3$ & $-13 \pm 1$ & D \\ 
 GJ 576 B & $-89.8 \pm 0.2 \pm 5$ & $-84.8 \pm 0.1$ & (3) & $-50 \pm 3$ & $-63.1 \pm 0.3$ & $-54 \pm 4$ & TD \\ 
 Gl 584 C & $-13.8 \pm 0.2 \pm 5$ & $-7.2 \pm 0.2$ & (9) & $22 \pm 2$ & $4 \pm 2$ & $-10 \pm 4$ & D \\ 
J1541$-$2250 & $-38.7 \pm 0.4 \pm 5$ & \nodata & \nodata & $-34 \pm 4$ & $2 \pm 1$ & $4 \pm 2$ & D \\ 
J1553$+$6933 & $-162.7 \pm 0.4 \pm 5$ & $110 \pm 90$\tablenotemark{b} & (1) & $-171 \pm 43$ & $-209 \pm 21$ & $-42 \pm 12$ & Halo \\ 
J1626$+$3925 & $-256.6^{+0.9}_{-1.0} \pm 5$ & $-239 \pm 12$\tablenotemark{b} & (6) & $-163 \pm 2$ & $-264 \pm 3$ & $-26 \pm 4$ & Halo \\ 
J1810$-$1010 & $-85.2^{+0.9}_{-1.1} \pm 5$ & $-85 \pm 13$\tablenotemark{b} & (11) & $-62 \pm 5$ & $-45 \pm 3$ & $34 \pm 2$ & D/TD \\ 
J1828$+$2650 & $-30.9 \pm 0.2 \pm 5$ & \nodata & \nodata & $-20 \pm 3$ & $12 \pm 4$ & $-41 \pm 2$ & D \\ 
 Wolf 1130 C & $-37.8 \pm 0.3 \pm 5$ & $-33 \pm 1$ & (7) & $117.6 \pm 0.1$ & $-36 \pm 5$ & $40 \pm 1$ & TD \\ 
J2030$+$0749 & $-28.3 \pm 0.1 \pm 5$ & $-21.2 \pm 0.7$ & (4) & $-20 \pm 3$ & $-9 \pm 4$ & $-11 \pm 2$ & D \\ 
 HN Peg B & $-19.8 \pm 0.2 \pm 5$ & $-20 \pm 1$ & (4) & $-4 \pm 2$ & $-11 \pm 4$ & $-2 \pm 2$ & D \\ 
\hline
\multicolumn{8}{c}{Additional Literature Metal-poor Brown Dwarf} \\
\hline
J1534$-$1043 & \nodata & $-116 \pm 5$ & (12) & $-66 \pm 4$ & $-181 \pm 17$ & $-110 \pm 5$ & Halo \\ 
\enddata
\tablerefs{
(1) \cite{Burgasser:2025aa}; (2) \cite{Gaia-Collaboration:2023aa}; (3) \cite{Halbwachs:2018aa}; (4) \cite{Hsu:2021aa}; (5) This work; (6) \cite{Lodieu:2015ab}; (7) \cite{Mace:2018aa}; (8) \cite{Reiners:2006ac}; (9) \cite{Soubiran:2018aa}; (10) \cite{Zhang:2018aa}; (11) \cite{Zhang:2025ad}; (12) \cite{Faherty:2025aa}.
}
\tablenotetext{a}{The RV uncertainties are reported in the order of sampling uncertainty and the systematic uncertainty of 5~{\kms}.}
\tablenotetext{b}{Measurement based on moderate-resolution data ($3,000 \lesssim \lambda/\Delta\lambda \lesssim 6,000$), which were ignored in our validation analysis.}
\tablenotetext{c}{Galactic kinematic populations, including thin disk (D), thick disk (TD), and halo (H). See Section~\ref{subsec:uvw} for the full definition.}
\tablenotetext{d}{While J1316$+$0755 shows halo kinematics using the \cite{Bensby:2003aa} formulation, it is assigned as a transitional thick disk/halo member from our analysis of Galactic total energy and Galactic angular momentum (Section~\ref{subsec:etot_lz}).}
\end{deluxetable*}

\section{Ancient Brown Dwarf Kinematics} \label{sec:kinematics}

\subsection{UVW Velocities} \label{subsec:uvw}

Combining astrometry with our measured RVs, we computed the Galactic \textit{UVW} space motions of the sources in our sample following the prescription described in \cite{Hsu:2021aa}.
We used the \cite{Johnson:1987aa} formulation to derive \textit{UVW} space motions, as implemented in the \texttt{SMART} function \texttt{compute\_uvw\_velocity} \citep{Hsu:2021ab}.
The \cite{Johnson:1987aa} formulation is a right-handed spherical coordinate system, where $U$ is the velocity toward the Galactic center, $V$ is the velocity along the Galactic rotation, and $W$ is the velocity toward the Galactic north pole.
We applied a heliocentric correction to place our $UVW$ space motions at the Local Standard of Rest (LSR) using the solar LSR velocity ($U, V, W$) = (11.1, 12.24, 7.25)~{\kms} from \cite{Schonrich:2010aa}.

From these velocities, we computed Galactic population probabilities using the velocity dispersions introduced in \cite{Bensby:2003aa}.
Galactic population probabilities of thick disk to thin disk probability ratio $P$(TD)/$P$(D) and halo to thick disk probability $P$(Halo)/$P$(TD) are computed using the \texttt{SMART} function ``compute\_galactic\_pop\_prob''.
Thin-disk membership is assigned for $P$(TD)/$P$(D) $< 0.1$; thick-disk membership is assigned for $P$(TD)/$P$(D) $> 10$; intermediate thin-disk/thick-disk membership is assigned for $0.1 <$ $P$(TD)/$P$(D) $< 10$.
Halo membership is assigned for $P$(Halo)/$P$(TD) $> 10$, and intermediate thick-disk/halo membership is assigned for $0.1 <$ $P$(Halo)/$P$(TD) $< 10$.
The derived velocities and kinematic population memberships 
for our sample are summarized in Table~\ref{tab:rv_uvw}.
We note that both J0414$-$5854 and J1316$+$0755 lack trigonometric parallaxes, so velocities and kinematic memberships are based on estimated distances from \cite{Schneider:2020aa} and \cite{Burningham:2014aa}, respectively.

We further evaluated the \textit{UVW} space motions of our sample by comparing them to previously identified stellar populations.
In particular, we considered the Galactic thin disk stars curated from the Gaia Catalogue of Nearby Stars (GCNS) \cite{Gaia-Collaboration:2021aa}, 
thick disk stars from \cite{Duong:2018aa}, 
halo stars from \cite{Nissen:2024aa}, 
Helmi stream members from \cite{Koppelman:2019aa} and \cite{Limbach:2021aa}, 
Thamnos stream members from \cite{Xie:2026aa}, 
and members of the Gaia Enceladus (GSE) dwarf galaxy merger from \cite{Ernandes:2024aa}.
For GCNS, we restricted the GCNS sample to RV errors less than 0.5~{\kms} and RUWE $<$ 1.3, and selected sources with thin disk memberships based on their Galactic kinematic membership probability (see Section~\ref{subsec:uvw}) using the same method in \cite{Bensby:2003aa}.
Where necessary, we updated the astrometry and RVs for these sources using data from 
Gaia DR3 \citep{Gaia-Collaboration:2023aa},
and computed LSR $UVW$ velocities in the same manner as our brown dwarf sample.

\begin{figure*}[ht!]
    \centering
    \includegraphics[trim=0 0cm 0 0cm, width=0.45\textwidth]{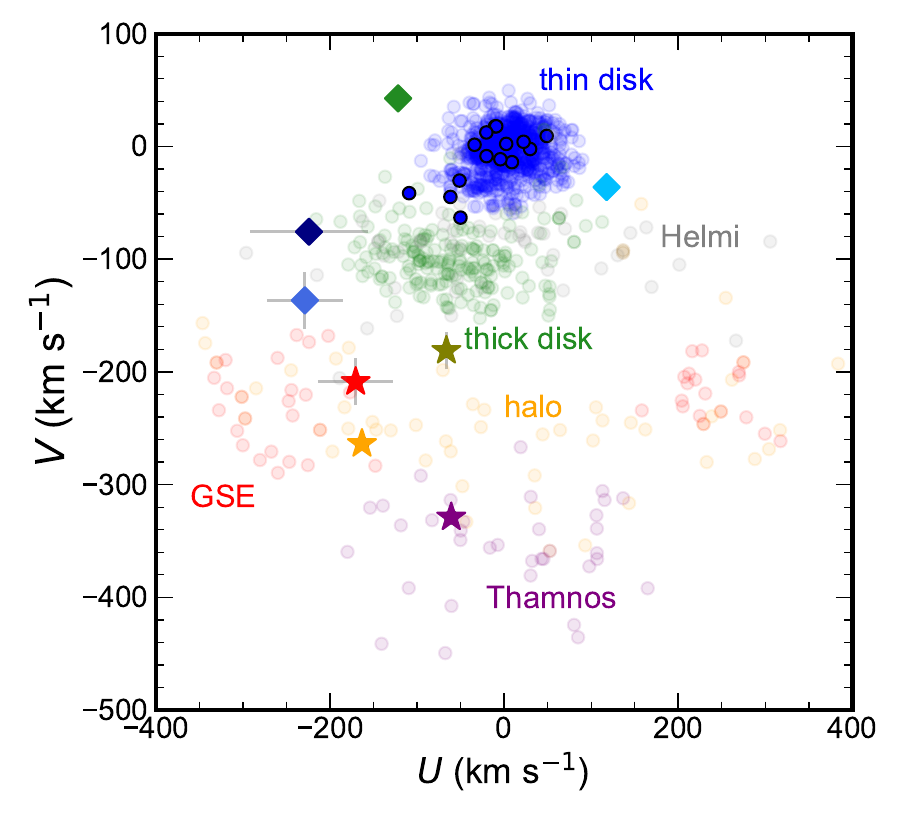}
    \includegraphics[trim=0 0cm 0 0cm, width=0.45\textwidth]{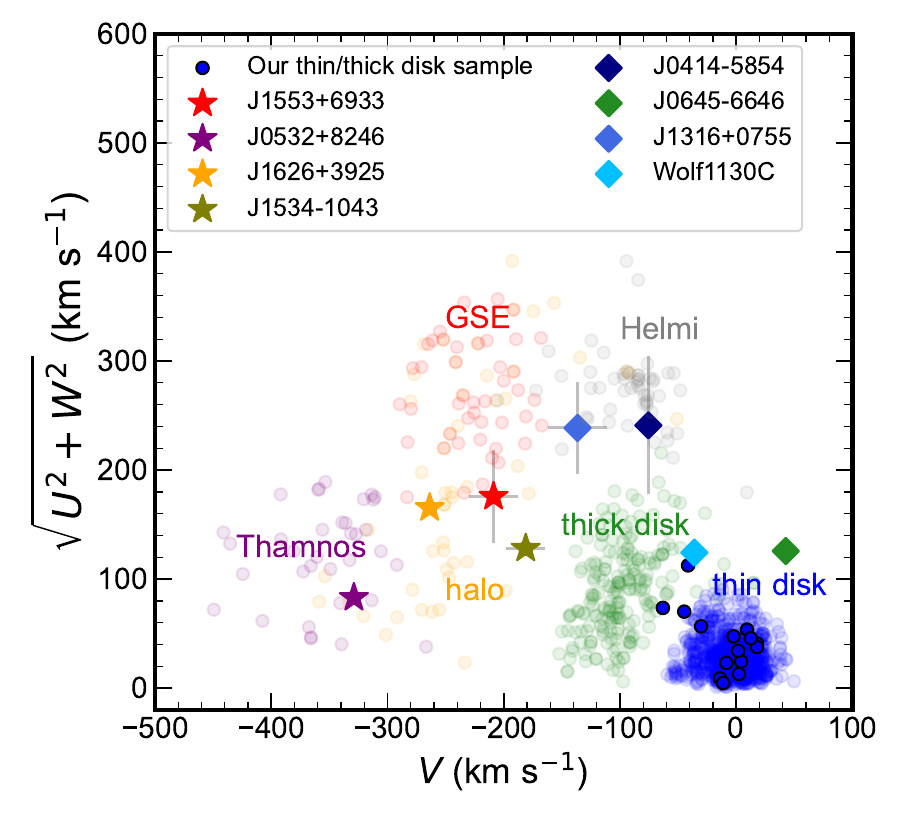}
    \includegraphics[trim=0 0cm 0 0cm, width=0.45\textwidth]{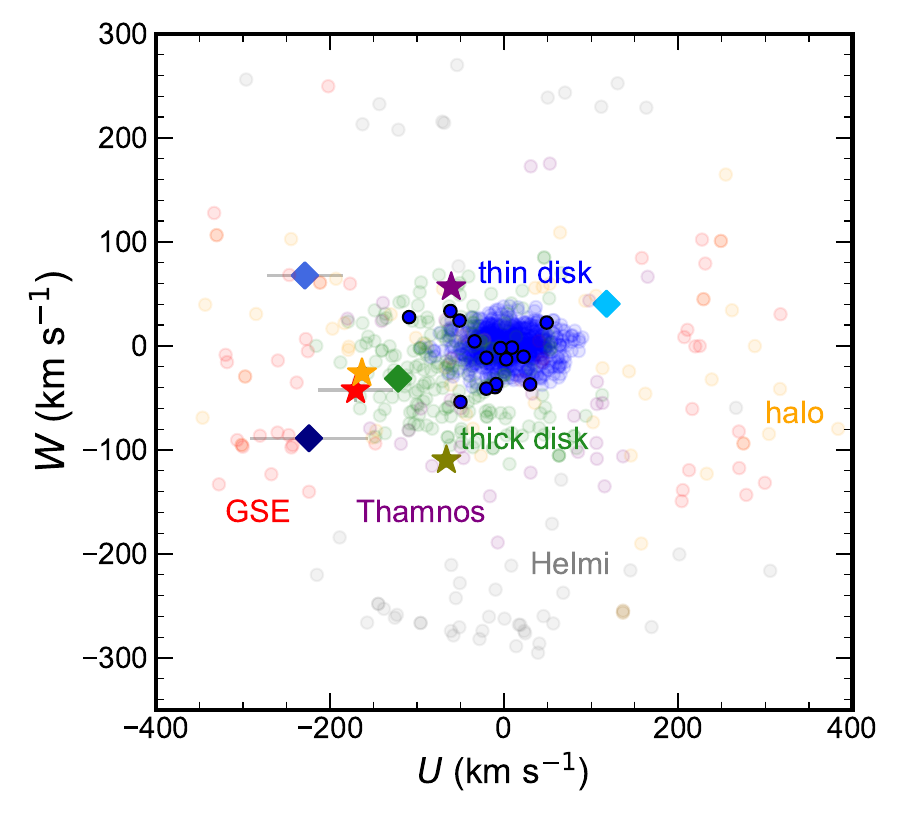}
    \includegraphics[trim=0 0cm 0 0cm, width=0.45\textwidth]{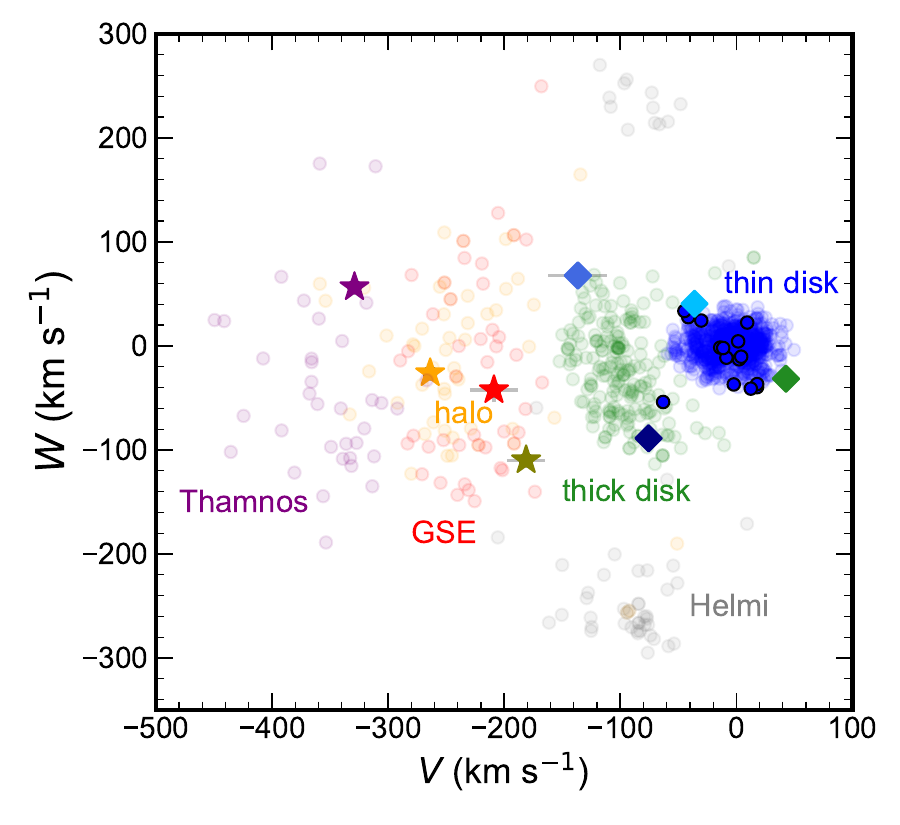}
    \caption{$UVW$ space motions and Toomre diagram for our sample.
    Individual sources indicated in the legend are labeled as stars (halo members) or diamonds (thick disk/halo members), whereas the remaining sources in our sample are shown as {large} blue dots {with black outlines}. Uncertainties are plotted, but are generally smaller than the marker size.
    $UVW$ velocities for stars in labeled Galactic populations and substructures are also shown, including 
    thin disk stars from \citet[][blue]{Gaia-Collaboration:2021aa}, 
    thick disk stars from \citet[][green]{Duong:2018aa}, 
    halo stars from \cite[][orange]{Nissen:2024aa}, 
    Helmi stream members from \cite{Koppelman:2019aa} and \citet[][grey]{Limbach:2021aa}, 
    Thamnos members from \citet[][purple]{Xie:2026aa}, 
    and Gaia Enceladus (GSE) members from \citet[][red]{Ernandes:2024aa}. 
    }\label{fig:uvw}
\end{figure*}

Figure~\ref{fig:uvw} illustrates the \textit{UVW} space motions of our brown dwarfs and the literature samples.
The majority of our sample (12 out of 23 sources) are aligned with the Galactic thin disk, in particular our solar-metallicity benchmark companion sources.
Our transitional thin disk/thick disk sources (J0448$-$1935 and J1810$-$1010) and thick disk sources (J0645$-$6646, GJ 576 B, and Wolf 1130 C) 
clump near the thin disk locus but extend outward toward higher velocities.
Our transitional thick disk/halo sources (J0414$-$5854 and J1316$+$0755\footnote{J1316$+$0755 shows halo kinematics from \cite{Bensby:2003aa} metrics, but in Section~\ref{subsec:etot_lz} we show that it is most consistent with thick disk membership, so it is designated as thick-disk/halo.}) both appear to overlap with Helmi stream members in the Toomre diagram and marginally in the \textit{U}-\textit{V} plane.
However, they are highly inconsistent with the Helmi stream
in the \textit{U}-\textit{W} and \textit{V}-\textit{W} planes.
Additionally, we note that both J0414$-$5854 and J1316$+$0755 lack trigonometric parallaxes, so more robust parallaxes could help {anchor} their true kinematic membership and origins.
Our halo sources are widely scattered in each of the \textit{UVW} planes,
with J1553$+$6933 overlapping with Gaia Enceladus members (cf. \citealp{Meisner:2026aa}), 
J0532$+$8246 overlapping with Thamnos members (cf.\ \citealt{Burgasser:2025aa}), 
and J1626$+$3925 and J1534$-$1043 appearing to be general halo members.
These potential memberships are evaluated further
in Section~\ref{subsec:etot_lz}.

\subsection{Galactic Orbits} \label{subsec:gal_orbit}

We assessed the Galactic orbits for our sample using \texttt{galpy} \citep{Bovy:2015aa}, following the approach described in \cite{Hsu:2021aa, Hsu:2024aa}.
We adopted the isocylindrical Galactic potential model of {\cite{McMillan:2017aa}}\footnote{{The Miyamoto-Nagai potential was used in several brown dwarf kinematics studies \citep{Burgasser:2015ac, Hsu:2021aa, Hsu:2024aa}.
To be fully consistent with our analysis in Section~\ref{subsec:etot_lz}, we use the \cite{McMillan:2017aa} potential, which also includes a halo component.
}}, 
assuming a solar azimuthal velocity of {233.1~{\kms}}, a solar Galactic radius and vertical height ($R_{\odot}$, $Z_{\odot}$) = {(8.21, 0.014) kpc \citep{McMillan:2017aa}}
and the LSR solar motion from \cite{Schonrich:2010aa} used to compute
brown dwarf $UVW$ velocities.
Orbit integrations spanned $-5$~Gyr to $+5$~Gyr relative to the present with a time step of 2~Myr.
We parameterized the resulting orbits through their 
minimum and maximum Galactic cylindrical radius ($R_\text{max}, R_\text{min}$), 
maximum absolute Galactic vertical height ($\left| Z \right|$), 
median orbit eccentricity ($e \equiv \langle R_\text{max} - R_\text{min} \rangle / \langle R_\text{max} + R_\text{min} \rangle$), 
and median absolute orbit inclination ($\tan{i} \equiv |Z / \sqrt{X^2 + Y^2}|$), where $XYZ$ are Galactic Cartesian coordinates aligned with our $UVW$ velocity components.
Uncertainties in astrometry and RV measurements were propagated using 100 Monte Carlo realizations of the Galactic orbits for each source. 

The inferred Galactic orbital parameters are summarized in Table~\ref{tab:gal_orbit}.
The majority of our thin disk sources have flat ($\left| Z \right|$ $\lesssim$ {300}~pc, $i$ $\lesssim$ 2$^{\circ}$), 
near-circular orbits ($R_{min} \approx R_\odot \gtrsim$ 7~kpc, $R_{max} \approx R_\odot \lesssim$ 10~kpc, $e$ $\lesssim$ 0.15), consistent with other local brown dwarfs \citep{Burgasser:2015ac, Hsu:2021aa, Hsu:2024aa}.
The one exception is the HD~126053 subdwarf system which has $\left| Z \right|$ = {670}~pc ($i$ = {4.5}$^{\circ}$) despite having a low radial eccentricity.
Compared to these sources, our thick disk and halo objects have progressively larger 
orbital eccentricities (median values: thin disk = {0.09}, thick disk = {0.38}, halo = {0.81})
and inclinations (thin disk = {1.9}$^{\circ}$, thick disk = {6}$^{\circ}$, halo = {27}$^{\circ}$), 
consistent with scattering from or formation outside the thin disk.
The radial motions of the thick disk and halo objects naturally divide into {two} categories:
``in-bound'' ($R_{min} \lesssim$ 7~kpc, $R_{max} \approx R_\odot \lesssim$ 10~kpc)
and ``passing through'' ($R_{min} \lesssim$ 7~kpc, $R_{max} \gtrsim$ 10~kpc).
Notably, the three sources with retrograde orbits ($L_z < 0$, see below) are largely on in-bound orbits, with 
J1553+6933 and J1626+3925 diving deeply into the bulge ($R <$ 1~kpc) on near-radial trajectories ($e$ $\approx$ 0.9).
In contrast, the thick disk dwarf and {halo} dwarf J0645$-$6646 {and J1553+6933} have extreme out-bound orbits extending 
to {15--20}~kpc from the Galactic center (cf.\ \citealt{Cushing:2009aa}).

\begin{deluxetable*}{lcccccccc}
\tablewidth{700pt}
\tablecaption{Galactic Orbital Parameters \label{tab:gal_orbit}} 
\tabletypesize{\scriptsize} 
\tablehead{ 
\colhead{Name} & 
\colhead{Pop.\tablenotemark{a}} & 
\colhead{$R_\text{min}$} & 
\colhead{$R_\text{max}$} & 
\colhead{$\left| Z \right|$} & 
\colhead{$e$} & 
\colhead{$i$} & 
\colhead{$E_\mathrm{tot}$} & 
\colhead{$L_z$} \\
\colhead{} & \colhead{} & \colhead{(kpc)} & \colhead{(kpc)} & \colhead{(kpc)} & \colhead{} & 
\colhead{(deg)} & \colhead{(km$^2$ s$^{-2}$)} & \colhead{(kpc {\kms})}
} 
\startdata
\hline
J0414$-$5854 & TD/Halo & $4.22^{+0.14}_{-0.08}$ & $8.5^{+0.2}_{-0.1}$ & $1.96^{+0.27}_{-0.09}$ & $0.34^{+0.01}_{-0.02}$ & $17.0^{+1.6}_{-0.5}$ & $-168456^{+568}_{-573}$ & $1304 \pm 27$ \\ 
J0415$-$0935 & D & $5.9 \pm 0.1$ & $8.79^{+0.06}_{-0.05}$ & $0.42^{+0.05}_{-0.07}$ & $0.19 \pm 0.01$ & $3.1^{+0.4}_{-0.5}$ & $-162012^{+234}_{-214}$ & $1724 \pm 13$ \\ 
J0448$-$1935 & D/TD & $4.9^{+0.1}_{-0.13}$ & $10.3 \pm 0.1$ & $0.6 \pm 0.1$ & $0.36 \pm 0.01$ & $4.0^{+0.7}_{-0.9}$ & $-159465^{+427}_{-376}$ & $1637^{+24}_{-20}$ \\ 
J0532$+$8246 & Halo & $2.2 \pm 0.1$ & $8.48^{+0.05}_{-0.06}$ & $1.57^{+0.07}_{-0.09}$ & $0.59 \pm 0.02$ & $19^{+1}_{-2}$ & $-175975^{+434}_{-453}$ & $-742^{+42}_{-27}$ \\ 
J0645$-$6646 & TD & $6.76 \pm 0.09$ & $15.9^{+0.8}_{-0.6}$ & $0.9 \pm 0.1$ & $0.41 \pm 0.02$ & $3.7^{+0.5}_{-0.4}$ & $-138271^{+1503}_{-1691}$ & $2326^{+32}_{-39}$ \\ 
 HIP 38939 B & D & $7.29^{+0.04}_{-0.07}$ & $10.1^{+0.2}_{-0.3}$ & $0.411^{+0.005}_{-0.006}$ & $0.162^{+0.009}_{-0.013}$ & $2.52^{+0.04}_{-0.02}$ & $-153201^{+1228}_{-1125}$ & $2047^{+37}_{-35}$ \\ 
J0836$-$1859 & D & $7.2^{+0.2}_{-0.3}$ & $8.21^{+0.05}_{-0.04}$ & $0.031^{+0.011}_{-0.005}$ & $0.068^{+0.016}_{-0.008}$ & $0.23^{+0.08}_{-0.03}$ & $-159968^{+898}_{-1107}$ & $1860^{+32}_{-41}$ \\ 
J1316$+$0755 & TD/Halo & $5.8 \pm 0.3$ & $8.8^{+0.2}_{-0.1}$ & $0.81^{+0.12}_{-0.1}$ & $0.2 \pm 0.03$ & $5.9^{+0.9}_{-0.8}$ & $-162296^{+544}_{-568}$ & $1680^{+41}_{-42}$ \\ 
J1416$+$1348 B & D & $8.087^{+0.014}_{-0.009}$ & $9.72^{+0.05}_{-0.06}$ & $0.77^{+0.09}_{-0.13}$ & $0.092^{+0.003}_{-0.004}$ & $4.7^{+0.5}_{-0.8}$ & $-151948^{+184}_{-175}$ & $2113.0^{+1.1}_{-0.9}$ \\ 
J1416$+$1348 A & D & $8.096^{+0.012}_{-0.01}$ & $9.69^{+0.04}_{-0.06}$ & $0.7^{+0.09}_{-0.13}$ & $0.089^{+0.003}_{-0.004}$ & $4.3^{+0.5}_{-0.7}$ & $-152055^{+136}_{-131}$ & $2113.7^{+0.9}_{-0.8}$ \\ 
 HD 126053 B & D & $7.39^{+0.07}_{-0.1}$ & $8.96^{+0.04}_{-0.02}$ & $0.67^{+0.07}_{-0.1}$ & $0.096^{+0.009}_{-0.006}$ & $4.5^{+0.5}_{-0.7}$ & $-156470^{+198}_{-208}$ & $1947^{+4}_{-5}$ \\ 
 HIP 70849 B & D & $8.06^{+0.04}_{-0.19}$ & $8.35^{+0.13}_{-0.05}$ & $0.17^{+0.01}_{-0.02}$ & $0.02^{+0.01}_{-0.003}$ & $1.2^{+0.09}_{-0.13}$ & $-156667^{+917}_{-559}$ & $1978^{+31}_{-20}$ \\ 
 GJ 576 B & TD & $4.704^{+0.003}_{-0.002}$ & $8.52^{+0.04}_{-0.06}$ & $1.16^{+0.1}_{-0.14}$ & $0.289^{+0.002}_{-0.003}$ & $9.3^{+0.7}_{-1.1}$ & $-167227^{+346}_{-306}$ & $1448 \pm 2$ \\ 
 Gl 584 C & D & $7.688^{+0.008}_{-0.011}$ & $8.9^{+0.2}_{-0.1}$ & $0.15^{+0.04}_{-0.07}$ & $0.075^{+0.009}_{-0.006}$ & $1.0^{+0.3}_{-0.5}$ & $-155837^{+539}_{-390}$ & $1999^{+18}_{-14}$ \\ 
J1541$-$2250 & D & $7.44^{+0.07}_{-0.05}$ & $9.1^{+0.1}_{-0.2}$ & $0.07^{+0.03}_{-0.02}$ & $0.099^{+0.01}_{-0.016}$ & $0.5 \pm 0.2$ & $-156185^{+396}_{-406}$ & $1980^{+9}_{-10}$ \\ 
J1553$+$6933 & Halo & $0.4^{+0.3}_{-0.1}$ & $17^{+3}_{-2}$ & $1.4^{+4.2}_{-1.0}$ & $0.957^{+0.009}_{-0.018}$ & $26^{+36}_{-19}$ & $-147378^{+11762}_{-8556}$ & $-132^{+142}_{-135}$ \\ 
J1626$+$3925 & Halo & $0.53^{+0.06}_{-0.05}$ & $10.44^{+0.08}_{-0.06}$ & $1.2^{+3.1}_{-0.3}$ & $0.904^{+0.008}_{-0.009}$ & $29^{+11}_{-4}$ & $-170291^{+487}_{-477}$ & $-196^{+28}_{-27}$ \\ 
J1810$-$1010 & D/TD & $5.3 \pm 0.1$ & $8.84^{+0.08}_{-0.09}$ & $0.64^{+0.05}_{-0.06}$ & $0.26^{+0.01}_{-0.02}$ & $4.9^{+0.4}_{-0.5}$ & $-164119^{+307}_{-274}$ & $1601^{+16}_{-22}$ \\ 
J1828$+$2650 & D & $7.95^{+0.07}_{-0.06}$ & $9.5 \pm 0.2$ & $0.79^{+0.04}_{-0.03}$ & $0.088^{+0.009}_{-0.005}$ & $5.0 \pm 0.3$ & $-153187^{+1012}_{-788}$ & $2065^{+39}_{-30}$ \\ 
 Wolf 1130 C & TD & $5.0^{+0.2}_{-0.1}$ & $11.0^{+0.2}_{-0.1}$ & $0.98^{+0.05}_{-0.03}$ & $0.375^{+0.005}_{-0.007}$ & $6.32^{+0.09}_{-0.05}$ & $-156603^{+920}_{-1221}$ & $1683^{+36}_{-48}$ \\ 
J2030$+$0749 & D & $7.3^{+0.3}_{-0.2}$ & $8.385^{+0.008}_{-0.004}$ & $0.15^{+0.03}_{-0.02}$ & $0.07^{+0.01}_{-0.02}$ & $1.1^{+0.2}_{-0.1}$ & $-158741^{+831}_{-859}$ & $1899^{+31}_{-33}$ \\ 
 HN Peg B & D & $7.4^{+0.3}_{-0.2}$ & $8.123^{+0.004}_{-0.005}$ & $0.03^{+0.03}_{-0.01}$ & $0.05^{+0.01}_{-0.02}$ & $0.19^{+0.2}_{-0.08}$ & $-159669^{+1001}_{-1042}$ & $1874^{+36}_{-39}$ \\ 
\hline
\multicolumn{9}{c}{Additional Literature Metal-poor Brown Dwarf} \\
\hline
J1534$-$1043 & Halo & $1.5^{+0.4}_{-0.5}$ & $8.8^{+0.2}_{-0.1}$ & $5.9^{+0.6}_{-1.1}$ & $0.71^{+0.08}_{-0.06}$ & $57^{+10}_{-8}$ & $-174846^{+1005}_{-897}$ & $495^{+129}_{-118}$ \\ 
\enddata
\tablenotetext{a}{Galactic kinematic populations, including thin disk (D), thick disk (TD), and halo (H). See Section~\ref{subsec:uvw} for the full definition.}
\end{deluxetable*}

\subsection{Galactic $E_\mathrm{tot}$-$L_z$ Analysis} \label{subsec:etot_lz}

While $UVW$ velocity components are convenient for assessing kinematic membership and tracing orbital motion, they are not conserved across the Galactic potential, making it difficult to compare stars in different regions across the Milky Way. For this reason,  
total orbital energy $E_\mathrm{tot}$ and vertical angular momentum $L_z$ are 
better parameters for identifying and assessing membership in Galactic populations and substructures \citep{Helmi:2020aa}.
We computed these quantities using \texttt{galpy}, assuming the \cite{McMillan:2017aa}\footnote{{The Galactic potential used to select the Galactic halo and substructures from \cite{Naidu:2020aa} is \cite{McMillan:2017aa}, and to compare their results with our sample, the same Galactic potential is necessary.}} Galactic potential, propagating uncertainties 
using 100 Monte Carlo samples.
In addition to our brown dwarfs, we also computed these quantities for the Galactic population samples described in Section~\ref{subsec:uvw}. 
To ensure robust membership for these samples, we limited thin disk, thick disk, and halo populations to those that satisfy the 
\cite{Bensby:2003aa} probability criteria (Section~\ref{subsec:uvw}), 
and constrained the Gaia Enceladus and Helmi samples to those sources with propagated uncertainties $\sigma (E_\mathrm{tot}) < 1000$~km$^2$ s$^{-2}$ and $\sigma (L_z) < 100$~kpc {\kms}.

\begin{figure*}[ht!]
    \centering
    \includegraphics[trim=1cm 0cm 0 0cm, width=0.9\textwidth]{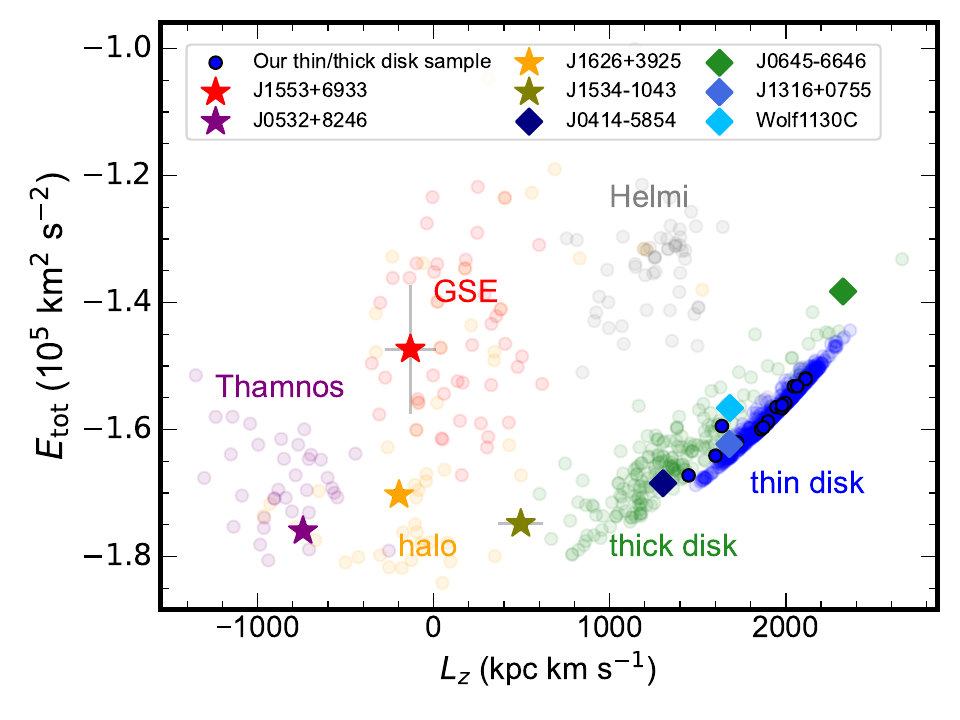}
    \caption{Galactic total energy ($E_\mathrm{tot}$) versus vertical angular momentum ($L_z$) for our sample, based on  
    \texttt{galpy} \citep{Bovy:2015aa} and the \texttt{McMillan17} Galactic potential \citep{McMillan:2017aa}.
    Symbols and colors are the same as Figure~\ref{fig:uvw}.
    }\label{fig:etotal_lz}
\end{figure*}

Figure~\ref{fig:etotal_lz} illustrates the $E_\mathrm{tot}$-$L_z$ space for these sources.
As expected, our solar or near-solar metallicity brown dwarfs are fully consistent with the thin disk population from Gaia, while 
our two transitional thin/thick disk sources, J0448$-$1935 and J1810$-$1010, lie in an overlapping space between the thin disk and thick disk samples.
Two of our thick disk sources, Wolf 1130 C and GJ 576 B, are in line with the bulk of the thick-disk population,
whereas J0645$-$6646 resides at the high energy, high angular momentum tail of this distribution.
Among the brown dwarfs exhibiting halo kinematics, J1553$+$6933 aligns well in $E_\mathrm{tot}$-$L_z$ space with Gaia Enceladus members, as also reported in \cite{Meisner:2026aa}; it can be firmly ruled out as a Helmi member as proposed by \citet{Burgasser:2025aa}.
We also confirm the L subdwarf J0532$+$8246 as a likely member of Thamnos based on these quantities proposed in \cite{Burgasser:2025aa}.
J1626$+$3925 and J1534$-$1043, on the other hand, are consistent with the Milky Way's in-situ halo population.
Except for J1534$-$1043, the remaining three sources are notable for exhibiting retrograde Galactic orbits ($L_z < 0$).

The T subdwarfs J0414$-$5854 (thick-disk/halo) and J1316$+$0755 (thick-disk/halo) warrant special attention.
As discussed in Section~\ref{sec:kinematics}, both overlapped with Helmi stream members in the \textit{U}-\textit{V} plane and Toomre plot, but not in the 
\textit{U}-\textit{W} plane and \textit{V}-\textit{W} planes. 
Figure~\ref{fig:etotal_lz} shows that they are also 
highly inconsistent with the Helmi stream in the $E_\mathrm{tot}$-$L_z$ plane.
Instead, both objects {reside} within the broader thick disk population. 
{Lacking a robust parallax measurement, our kinematic population assignments are} preliminary for these two sources.

Verifying the true population membership of these sources will require characterization of their chemical abundances, analysis which will be presented in future studies (cf. \citealp{Meisner:2026aa} for J1553$+$6933).
Nevertheless, our analysis demonstrates the necessity of considering multiple kinematic diagnostics, in addition to chemical abundances, for
determining the Galactic population memberships of nearby sources when comparing to more distributed stellar samples.

\section{Individual Objects of Interest}\label{sec:individual}

In this Section, we discuss objects that exhibit thick disk or halo kinematics and summarize their {relevant} literature together with our findings in this work, including J0414$-$5854, J0532$+$8246, J0645$-$6646, J1316$+$0755, J1553$+$6933, J1626$+$3925, J1810$-$1010, and Wolf 1130 C.
Here, we only focus on the fundamental and kinematic properties of metal-poor brown dwarfs in our sample.
Some of the parameters quoted in this section may change significantly {in the upcoming years}, as the metal-poor self-consistent models are relatively new {owing} to their exotic chemistry and rapidly updated opacities.
For observed properties such as color, spectral features, and classifications, we refer readers to \cite{Burgasser:2025aa} and other earlier works (e.g., \citealp{Lodieu:2015ab, Zhang:2017aa, Zhang:2019aa, Lodieu:2019ab, Schneider:2020aa, Meisner:2020aa, Meisner:2021aa}{, and references in Tables~\ref{tab:sample} and \ref{tab:observations}}).

\textit{J0414$-$5854}: WISE J041451.64$-$585457.0 was discovered in \cite{Schneider:2020aa},
as one of the prototype T-type extreme subdwarfs.
It is an extreme subdwarf (esd) T6 with {\teff} = $699^{+66}_{-49}$~K and [M/H] = $-1.34^{+0.17}_{-0.15}$~dex in \cite{Burgasser:2025aa}.
Using our updated proper motions and RV, it is a transitional thick-disk/halo member.
From our kinematic analysis, it exhibits a {moderate} Galactic eccentricity of {$0.34^{+0.01}_{-0.02}$} and inclination of {$17.0^{+1.6}_{-0.5}$}~deg, most likely to be a thick disk member and not associated with the Helmi Stream.
{However, t}he membership of this first T extreme subdwarf should be revisited once its parallax is available.

\textit{J0532$+$8246}: 2MASS J053253.46$+$824646.5 is an extreme L subdwarf, with a spectral type of esdL7 \citep{Burgasser:2003ac, Kirkpatrick:2010aa, Zhang:2017aa}.
It is a well-characterized metal-poor late L-type brown dwarf \citep{Kirkpatrick:2005aa, Burgasser:2007ac}.
While J0532$+$8246 is substellar, its optical spectra lack Lithium \citep{Lodieu:2015ab}, {a species }expected for objects {of} masses {below} 0.060~{\msun} \citep{Baraffe:2003aa}.
It has a [Fe/H] $\sim$ $-1.6$~dex and {\teff} of 1550$\pm$175~K \citep{Burgasser:2025aa}, and shows halo kinematics and a retrograde Galactic eccentric orbit \citep{Burgasser:2008ac} and consistent with the Thamnos Stream based on the $U$-$V$ space and Toomre analysis \citep{Burgasser:2025aa}.
Based on our RV measurement and kinematic analysis, we confirm that it is a likely member of Thamnos based on our full analysis of $UVW$ motions and its $E_\mathrm{tot}$-$L_z$.
We updated the Galactic orbits of $e = 0.59 \pm 0.02$ and constrained its high inclination of $i = 19^{+1}_{-2}$~deg.

\textit{J0645$-$6646}: 2MASS J06453153$-$6646120 is discovered \citep{Kirkpatrick:2010aa} and reclassified as d/sdT0 \citep{Burgasser:2025aa}.
Its optical spectra lack Lithium \citep{Lodieu:2015ab}, {with} its {\teff} = 1419$\pm$81~K and [M/H] $\sim$ $-0.4$ to $+0.8$~dex \citep{Burgasser:2025aa}.
We present its RV measurement (consistent with RV = $-33 \pm 10$~{\kms} using the VLT X-shooter spectra from \citealp{Zhang:2018aa}), with clear thick-disk kinematics.
Interestingly, its eccentric ($e = 0.41 \pm 0.02$) Galactic orbit is highly prograde and slightly deviates from the bulk thick disk population, and its maximum Galactic radius of $15.9^{+0.8}_{-0.6}$~kpc is significantly larger than the majority of our sample, other than {J1553$+$6933}.
It was flagged as a potential binary but with low probability \citep{Brooks:2023ab}. While we observed slight mismatches of the spectral fit, in 4.66--4.72~{\micron}, the mismatch is not significant enough to support {the} hypothesis {that it is a binary} compared to known short-period double-line ultracool dwarf binaries \citep{Hsu:2023aa}.

\textit{J1316$+$0755}: J131609.68+075553.9 is a sdT6.5 dwarf discovered in \cite{Burningham:2014aa}.
J1316$+$0755 has a {\teff} of 765$^{+55}_{-50}$~K and [M/H] = $-0.65^{+0.11}_{-0.09}$~dex \citep{Burgasser:2025aa}.
It also has {a moderately} eccentric and inclined Galactic orbit ($e = 0.20 \pm 0.03$ and $i = 5.9^{+0.9}_{-0.8}$~deg.
Our kinematic analysis indicates that it is a transitional thick-disk and halo object.
The evidence of halo membership is based on its $UVW$ and Toomre positions and \cite{Bensby:2003aa} metrics (Figure~\ref{fig:uvw}), while our analysis of $E_\mathrm{tot}$-$L_z$ places it confidently as a thick disk member (Figure~\ref{fig:etotal_lz}).
Similar to our discussion of J0414$-$5854, we rule J1316$+$0755 out as a member {of} the Helmi Stream.
{However, }its kinematics should be revisited once its parallax is available.

\textit{J1553$+$6933}: This brown dwarf was discovered in \cite{Meisner:2020aa}
as a sdT4 dwarf with metallicity $\sim$ $-0.5$~dex.
Recent follow-up analysis indicates that J1553$+$6933 {has} {\teff} = 1225$^{+70}_{-75}$~K and [M/H] = $-1.0^{+0.2}_{-0.3}$~dex, and previously it was proposed as a member of the Helmi Stream \citep{Burgasser:2025aa}.
Using our updated RV, J1553$+$6933 is a confirmed halo object associated with the Gaia Enceladus.
\cite{Meisner:2026aa} found J1553$+$6933 has effective temperature {\teff} = 686--990~K, metallicity [M/H] = $-$1.25$^{+0.14}_{-0.21}$~dex and $\alpha$-enrichment [$\alpha$/Fe] = $+$0.16$^{+0.03}_{-0.04}$~dex.
We further verified that all kinematic properties are consistent with the properties defined in the sample curated in \cite{Ernandes:2024aa}.
{Our kinematic analyses based on} $E_\mathrm{tot}$-$L_z$ {and} Galactic orbital eccentricity ($e = 0.957^{+0.009}_{-0.018}$) {make} its kinematic association with Gaia Enceladus highly confident.
Its Galactic high orbital inclination ($i = 26^{+36}_{-19}$~deg), small inner Galactic radius ($0.4^{+0.3}_{-0.1}$~kpc), and retrograde Galactic motion are reminiscent of its halo kinematics.
{It is noted that J1553$+$6933 does not exhibit any evidence of being a close binary.}

\textit{J1626$+$3925}: 2MASS J16262034$+$3925190 was discovered as one of the first L-type subdwarfs and pointed out as a halo object \citep{Burgasser:2004ac}, and classified as esdL4 \citep{Burgasser:2007ac, Gizis:2006aa, Zhang:2013aa, Zhang:2017aa}.
It is likely a metal-poor low-mass star \citep{Burgasser:2004ac, Schilbach:2009aa}, consistent with the non-detection of Lithium \citep{Lodieu:2015ab}.
It has {\teff} $= 2148 \pm 14$~K and [M/H] $\sim$ $-1.5$ to $-2.0$~dex \citep{Burgasser:2007ac, Burgasser:2009ac, Gonzales:2018aa}, and it has no known companions \citep{Goldman:2008aa}.
With our refined RV, all of our kinematic analyses place it as a halo object.
Its Galactic orbit is {retrograde, }highly eccentric ($e = 0.904^{+0.008}_{-0.009}$) and inclined ($i = 29^{+11}_{-4}$~deg).

\textit{J1810$-$1010}: WISEA J181006.18$-$101000.5 was discovered in \cite{Schneider:2020aa}, and classified as esdT3 \citep{Burgasser:2025aa}.
J1810$-$1010 is nearby ($\sim$9~pc; \citealp{Lodieu:2022aa}), with {\teff} = 835$\pm$58~K \citep{Burgasser:2025aa} and [Fe/H] = $-1.7 \pm 0.2$~dex \citep{Zhang:2025ad, Lodieu:2022aa}.
{Its {\teff} is }consistent with {the} detection of CH$_4$ \citep{Zhang:2025ad}.
Our companion paper (N. Lodieu et al. 2026, submitted) shows that J1810$-$1010 has a metallicity [M/H] = $-$1.40$^{+0.18}_{-0.11}$~dex and $\alpha$-enrichment [$\alpha$/Fe] = $+$0.15$^{+0.15}_{-0.03}$~dex.
{Kinematically, o}ur refined RV indicates that J1810$-$1010 is a transitional thin-/thick-disk object, with a moderately eccentric and inclined Galactic orbit ($e = 0.26^{+0.01}_{-0.02}$, $i = 4.9^{+0.4}_{-0.5}$~deg).

\textit{Wolf 1130 C}: Wolf 1130 C is known {for} an outlier with abundant PH$_3$ absorption \citep{Burgasser:2025ab}.
{D}iscovered in \cite{Mace:2013ab}, {Wolf 1130 C is a tertiary component comoving with} a close binary system {(i.e., in a triple system)}.
{The close binary is composed of} sdM1 GJ 781 A \citep{Gizis:1997aa} and white dwarf Wolf 1130B %
in a 0.5~day orbit \citep{Mace:2018aa}.
Wolf 1130 C was originally classified as a sdT8 dwarf \citep{Mace:2013ab, Logsdon:2018aa}.
\cite{Burgasser:2025ab} {reclassified it as (e)sdT6 }
with a {\teff} = $621 \pm 9$~K and [M/H] = $-0.68 \pm 0.04$~dex.
Our kinematic analysis and RV place the Wolf 1130 system as a member of the typical thick disk population, with a moderately eccentric and inclined Galactic orbit.

\section{Summary}\label{sec:sum}

We have presented the first RV survey of brown dwarfs using JWST/NIRSpec G395H spectroscopy.
Our main findings are summarized as follows:

\begin{itemize}
    \item We have compiled a kinematic sample of 23 brown dwarfs spanning a broad range of temperatures and metallicities observed with JWST NIRSpec. We report new proper motion measurements for two T subdwarfs,
    J0414$-$5854 and J1316$+$0755, based on multi-epoch observations from WISE, VHS, NSC DR2, and UKIDSS, and new high-resolution RV measurements of J1416$+$1348A from Keck/NIRSPEC spectra.
    \item We have introduced a forward-modeling framework for NIRSpec G395H spectra focused on the 4.4--5.0~{\micron} region, sampling the CO fundamental band. We demonstrate a statistical precision of $<$0.5~{\kms} and a systematic accuracy of 5~{\kms} (1/20$^\mathrm{th}$ of a resolution element, or 1/10$^\mathrm{th}$ of a pixel), the {latter} based on {comparison with} previously reported RVs obtained with ground-based high-resolution spectroscopy.
    \item Combining astrometry 
    and RVs, we determined \textit{UVW} space motions for our complete sample to median precisions of 3~{\kms}.
    These measurements allow us to associate our targets with various Galactic populations.
    Half of our sample, primarily benchmark companions to solar/near-solar metallicity stars, are thin disk objects;
    J0448$-$1935 and J1810$-$1010 are transitional thin/thick disk objects; 
    J0645$-$6646, GJ 576 B, Wolf 1130 C are thick disk objects; 
    J0414$-$5854 and J1316$+$0755 are transitional thick disk/halo objects; 
    and J0532$+$8246, J1553$+$6933, J1626$+$3925, and J1534$-$1043 are all halo objects.
    \item We inferred Galactic orbits for our sample, finding a progressive increase in eccentricity and absolute inclination from thin disk to thick disk to halo. Sources in the last two populations naturally divide into radially in-bound and passing trajectories. The halo objects J1553$+$6933 and J1626$+$3925 have the most eccentric in-bound orbits, passing within 1~kpc of the Galactic center at perigalacticon;
    while out-bound subdwarf J0645$-$6646 {and J1553$+$6933} have orbits that take them to Galactic radii beyond {15}~kpc.
    \item Analyzing the total orbital energy and angular momentum of our sources,
    we have confirmed that J1553$+$6933 is likely to be a member of Gaia Enceladus, consistent with its elemental abundances; {and} J0532$+$8246 is an excellent match to the Thamnos Stream.
\end{itemize}

This study demonstrates that JWST/NIRSpec G395H spectroscopy provides robust and reliable RVs for cool and faint brown dwarfs 
sufficient for Galactic kinematic studies.
Combined with the ability to infer detailed abundances from these data through retrieval modeling 
\citep{Faherty:2024aa,Faherty:2025aa,Rowland:2024aa,Burgasser:2025ab},
G395H spectra enable brown dwarfs to be physically characterized in the same precise manner as hotter stars. 
By combining measurements from the 6D kinematics plus abundances of these local brown dwarfs with the physical properties and spatial distributions of more distant brown dwarfs in 
the Galactic thick disk and halo \citep{Aganze:2022aa, Aganze:2022ab,Burgasser:2024aa, Hainline:2026aa},
globular clusters \citep{Gerasimov:2022aa, Gerasimov:2024aa, Gerasimov:2024ab}, 
and even the SMC \citep{Zeidler:2024aa},
new insights can be gained about
the Milky Way's formation history and assembly from its lowest-mass members.
{When hundreds of brown dwarfs in the Galactic halo and substructures are confirmed, constraints on the substellar mass function and star formation rate across cosmic age and in metal-poor environments will be possible beyond local environments \citep{Kirkpatrick:2021aa, Kirkpatrick:2024aa, Best:2024ab}.}

\facilities{JWST (NIRSpec)}

\software{
\texttt{Astropy} \citep{Astropy-Collaboration:2013aa, Astropy-Collaboration:2018aa}, 
\texttt{corner} \citep{Foreman-Mackey:2016aa},
\texttt{emcee} \citep{Foreman-Mackey:2013aa},
\texttt{galpy} \citep{Bovy:2015aa},
\texttt{Matplotlib} \citep{Hunter:2007aa}, 
\texttt{Numpy} \citep{Harris:2020aa},
\texttt{PyAstronomy} \citep{Czesla:2019aa},
\texttt{Scipy} \citep{Virtanen:2020aa}, 
\texttt{SPLAT} \citep{Burgasser:2017ac}, 
\texttt{SMART} \citep{Hsu:2021aa, Hsu:2021ab}
          }

\begin{acknowledgments}
{The authors thank the anonymous referee for their useful suggestions that improved this manuscript.}
CCH, CAT, AJB, EG, and GS acknowledge funding support from NASA/STScI through
JWST general observer program GO-4668, under NASA contract NAS 5-03127.  
AJB and EG acknowledge funding support from the Heising-Simons Foundation.
NL acknowledge support from the Agencia Estatal de Investigaci\'on del Ministerio de Ciencia e Innovaci\'on (AEI-MCINN) under grant PID2022-137241NB-C41\@.
All of the data presented in this paper were obtained from the Mikulski Archive for Space Telescopes (MAST) at the Space Telescope Science Institute. The specific observations analyzed can be accessed via \dataset[https://doi.org/10.17909/zm7x-dx89]{https://doi.org/10.17909/zm7x-dx89}. STScI is operated by the Association of Universities for Research in Astronomy, Inc., under NASA contract NAS5–26555. Support to MAST for these data is provided by the NASA Office of Space Science via grant NAG5–7584 and by other grants and contracts.
\end{acknowledgments}

\clearpage

\appendix
\restartappendixnumbering 
\section{Best-fit Parameters for Our Sample}

In Section~\ref{sec:model}, we show our forward-model method. Here we provide the Sonora Elf-Owl best-fit substellar atmosphere model parameters for interested readers to reproduce our results.
Other than RV, our best-fit parameters listed here should not be used as accurate determinations, as several studies in the literature have shown that focusing on a relatively narrow wavelength range could bias the true parameters of ultracool dwarfs (e.g., \citealp{Hsu:2021aa, Hsu:2024aa, Hsu:2024ab}), and our forward-model includes a 2$^\mathrm{nd}$-order polynomial continuum correction which would bias the inferred {\teff} and other parameters.
As for {\vsini}, the resolution of our G395H spectra is $\sim$3300 at 4.4--5.0~{\micron}, which corresponds to a {\vsini} floor $\sim$90.8~{\kms}.
Most brown dwarfs have {\vsini} $<$90~{\kms} \citep{Hsu:2021aa, Tannock:2021aa, Hsu:2024aa, Hsu:2026aa}, and G395H resolution is insufficient to provide robust {\vsini} (see for example, discrepant {\vsini} of J1828$+$2650 was reported using G395H spectra in \citealp{Lew:2024aa}).

\begin{deluxetable*}{lccccccc}[h!]
\tablewidth{700pt}
\tablecaption{Sonora Elf-Owl Best-fit Parameters for Our Sample\label{tab:mcmc_fit_param}} 
\tabletypesize{\footnotesize} 
\tablehead{ 
\colhead{Name} & 
\colhead{{\teff}} & 
\colhead{{\logg}} & 
\colhead{[M/H]} & 
\colhead{C/O\tablenotemark{a}} & 
\colhead{$\log{K_{zz}}$} & 
\colhead{{\vsini}} & 
\colhead{RV} \\
\colhead{} & \colhead{(K)} & \colhead{(cm s$^{-2}$)} & \colhead{(dex)} & \colhead{} &
\colhead{(cm$^2$ s$^{-1}$)} & \colhead{({\kms})} & \colhead{({\kms})}
} 
\startdata
\hline
J0414$-$5854 & $818^{+4}_{-5}$ & $5.253^{+0.012}_{-0.009}$ & $-0.995^{+0.006}_{-0.003}$ & $0.24^{+0.02}_{-0.01}$ & $5.4^{+0.2}_{-0.4}$ & $27^{+22}_{-17}$ & $130.3^{+0.4}_{-0.5}$ \\ 
J0415$-$0935 & $699^{+2}_{-4}$ & $3.92 \pm 0.04$ & $-0.84 \pm 0.02$ & $0.39^{+0.02}_{-0.03}$ & $7.12^{+0.16}_{-0.09}$ & $16^{+9}_{-10}$ & $46.8 \pm 0.2$ \\ 
J0448$-$1935 & $922 \pm 5$ & $4.55^{+0.04}_{-0.03}$ & $-0.997^{+0.006}_{-0.002}$ & $0.2303^{+0.0026}_{-0.001}$ & $6.93 \pm 0.01$ & $53^{+5}_{-8}$ & $92.0 \pm 0.3$ \\ 
J0532$+$8246 & $1789 \pm 6$ & $5.496^{+0.003}_{-0.007}$ & $-0.994^{+0.009}_{-0.004}$ & $0.239^{+0.011}_{-0.008}$ & $7.7^{+0.1}_{-0.3}$ & $44^{+3}_{-6}$ & $-173.4 \pm 0.2$ \\ 
J0645$-$6646 & $1672 \pm 11$ & $5.49^{+0.01}_{-0.02}$ & $-0.36 \pm 0.04$ & $0.382 \pm 0.007$ & $7.8^{+0.1}_{-0.3}$ & $55^{+5}_{-9}$ & $-25.4 \pm 0.3$ \\ 
 HIP 38939 B & $1233^{+10}_{-9}$ & $5.03^{+0.07}_{-0.1}$ & $-0.55^{+0.03}_{-0.04}$ & $0.28 \pm 0.02$ & $7.2^{+0.3}_{-0.2}$ & $17^{+12}_{-11}$ & $-14.0 \pm 0.2$ \\ 
J0836$-$1859 & $600^{+10}_{-9}$ & $3.6^{+0.06}_{-0.07}$ & $-0.988^{+0.016}_{-0.009}$ & $0.29 \pm 0.03$ & $7.2 \pm 0.2$ & $24^{+14}_{-15}$ & $21.6^{+0.5}_{-0.6}$ \\ 
J1316$+$0755 & $606^{+5}_{-4}$ & $4.1^{+0.05}_{-0.04}$ & $-0.96 \pm 0.02$ & $0.232^{+0.004}_{-0.002}$ & $6.95^{+0.01}_{-0.02}$ & $59^{+4}_{-9}$ & $24.8 \pm 0.4$ \\ 
J1416$+$1348 B & $638 \pm 2$ & $4.61 \pm 0.02$ & $-0.691 \pm 0.005$ & $0.2301^{+0.0016}_{-0.0008}$ & $4.4 \pm 0.3$ & $24^{+6}_{-12}$ & $-51.5 \pm 0.2$ \\ 
J1416$+$1348 A & $1900 \pm 7$ & $5.44^{+0.03}_{-0.04}$ & $-0.65 \pm 0.03$ & $0.457^{+0.007}_{-0.005}$ & $7.97^{+0.03}_{-0.06}$ & $24^{+14}_{-15}$ & $-48.2 \pm 0.2$ \\ 
 HD 126053 B & $545^{+6}_{-9}$ & $3.9 \pm 0.04$ & $-0.989^{+0.01}_{-0.008}$ & $0.231^{+0.003}_{-0.002}$ & $6.924 \pm 0.008$ & $29^{+7}_{-15}$ & $-24.8 \pm 0.3$ \\ 
 HIP 70849 B & $1342 \pm 5$ & $5.49^{+0.007}_{-0.016}$ & $0.612^{+0.01}_{-0.011}$ & $0.419 \pm 0.005$ & $5.4^{+0.4}_{-0.5}$ & $25^{+9}_{-13}$ & $-4.9 \pm 0.2$ \\ 
 GJ 576 B & $978 \pm 5$ & $5.0 \pm 0.01$ & $-0.494^{+0.011}_{-0.005}$ & $0.32^{+0.01}_{-0.02}$ & $5.6^{+0.4}_{-0.6}$ & $33^{+7}_{-14}$ & $-89.8 \pm 0.2$ \\ 
 Gl 584 C & $1677 \pm 7$ & $5.48^{+0.02}_{-0.03}$ & $0.23 \pm 0.02$ & $0.618 \pm 0.007$ & $7.2 \pm 0.2$ & $9^{+9}_{-6}$ & $-13.8 \pm 0.2$ \\ 
J1534$-$1043 & $832^{+25}_{-32}$ & $5.486^{+0.01}_{-0.021}$ & $-0.98^{+0.02}_{-0.01}$ & $0.6^{+0.4}_{-0.2}$ & $6.9^{+0.6}_{-0.2}$ & $195^{+3}_{-6}$ & $-130 \pm 5$ \\ 
J1541$-$2250 & $475^{+4}_{-3}$ & $3.52^{+0.05}_{-0.03}$ & $-0.997^{+0.005}_{-0.002}$ & $0.55^{+0.07}_{-0.09}$ & $7.91^{+0.06}_{-0.09}$ & $10^{+9}_{-8}$ & $-38.7 \pm 0.4$ \\ 
J1553$+$6933 & $1011 \pm 5$ & $5.11 \pm 0.02$ & $-0.996^{+0.007}_{-0.003}$ & $0.233^{+0.006}_{-0.003}$ & $6.91 \pm 0.01$ & $48^{+9}_{-18}$ & $-162.7 \pm 0.4$ \\ 
J1626$+$3925 & $2398^{+1}_{-3}$ & $3.57^{+0.06}_{-0.08}$ & $-0.99^{+0.02}_{-0.01}$ & $0.56^{+0.02}_{-0.03}$ & $7.7^{+0.2}_{-0.4}$ & $48^{+14}_{-21}$ & $-256.6^{+0.9}_{-1.0}$ \\ 
J1810$-$1010 & $1137^{+11}_{-18}$ & $5.497^{+0.002}_{-0.012}$ & $-0.998^{+0.006}_{-0.002}$ & $0.237^{+0.036}_{-0.006}$ & $5.0^{+0.5}_{-1.0}$ & $119^{+43}_{-20}$ & $-85.2^{+0.9}_{-1.1}$ \\ 
J1828$+$2650 & $400.0 \pm 0.3$ & $4.252^{+0.013}_{-0.009}$ & $-0.634 \pm 0.007$ & $0.2297^{+0.0011}_{-0.0005}$ & $6.972 \pm 0.008$ & $3^{+4}_{-2}$ & $-30.9 \pm 0.2$ \\ 
 Wolf 1130 C & $665^{+5}_{-8}$ & $4.98^{+0.02}_{-0.07}$ & $-0.82^{+0.01}_{-0.03}$ & $0.236^{+0.01}_{-0.006}$ & $5.7^{+0.2}_{-0.3}$ & $16^{+16}_{-11}$ & $-37.8 \pm 0.3$ \\ 
J2030$+$0749 & $1494^{+6}_{-7}$ & $5.492^{+0.006}_{-0.013}$ & $0.34 \pm 0.01$ & $0.69^{+0.006}_{-0.003}$ & $6.52^{+0.08}_{-0.11}$ & $19^{+8}_{-11}$ & $-28.3 \pm 0.1$ \\ 
 HN Peg B & $1365^{+24}_{-21}$ & $5.48^{+0.01}_{-0.03}$ & $0.61 \pm 0.03$ & $0.492 \pm 0.009$ & $6.5^{+0.2}_{-0.5}$ & $22^{+7}_{-14}$ & $-19.8 \pm 0.2$ \\ 
\enddata
\tablenotetext{a}{The solar elemental abundance C/O = 0.458.}
\end{deluxetable*}

\clearpage

\bibliography{mylibrary}{}
\bibliographystyle{aasjournalv7.1}

\end{document}